\documentclass{emulateapj}
\def \emulmacro {}
\usepackage{apjfonts}
\usepackage{graphicx}
\usepackage{amsmath}
\newcommand{\heoneonelong}{\mbox{HE\,1104$-$1805}}
\newcommand{\heoneone}{\mbox{HE\,1104}}
\newcommand{\qtwotwolong}{\mbox{Q\,2237$+$0305}}
\newcommand{\qtwotwo}{\mbox{Q\,2237}}
\newcommand{\hefourlong}{\mbox{HE\,0435$-$1223}}
\newcommand{\hefour}{\mbox{HE\,0435}}
\newcommand{\rxoneonelong}{\mbox{RX\,J1131$-$1231}}

\newcommand{\qzeroonelong}{\mbox{Q J0158$-$4325}}
\newcommand{\err}[2]{\ensuremath{^{_{+#1}}_{^{-#2}}}}
\ifx \emulmacro \undefined
\else
\fi
\begin{document}
\bibliographystyle{apj}
\shorttitle{THE STRUCTURE OF HE\,1104$-$1805 FROM INFRARED TO X-RAY}
\shortauthors{BLACKBURNE ET AL.}
\slugcomment{Submitted to ApJ}
\title{The Structure of HE\,1104$-$1805 from Infrared to X-Ray\altaffilmark{1}}

\author{
  Jeffrey~A.~Blackburne\altaffilmark{2},
  Christopher~S.~Kochanek\altaffilmark{2,3},
  Bin~Chen\altaffilmark{4}, \&
  Xinyu~Dai\altaffilmark{4}
  George~Chartas\altaffilmark{5},
}

\altaffiltext{1}{Based on observations made with the NASA/ESA Hubble
  Space Telescope, obtained at the Space Telescope Science Institute,
  which is operated by the Association of Universities for Research in
  Astronomy, Inc., under NASA contract NAS 5-26555. These observations
  are associated with programs \#11732 and \#12324.}
\altaffiltext{2}{Department of Astronomy, The Ohio State University,
  140 West 18th Avenue, Columbus, OH 43210, USA;
  blackburne@astronomy.ohio-state.edu}
\altaffiltext{3}{Center for Cosmology and AstroParticle Physics, The
  Ohio State University, Columbus, OH 43210, USA}
\altaffiltext{4}{Homer L. Dodge Department of Physics and Astronomy,
  University of Oklahoma, Norman, OK 73019, USA}
\altaffiltext{5}{Department of Physics and Astronomy, College of
  Charleston, Charleston, SC 29424, USA}
\addtocounter{footnote}{5}
\begin{abstract}

The gravitationally lensed quasar \heoneonelong\ has been observed at
a variety of wavelengths ranging from the mid-infrared to X-ray for
nearly 20 years. We combine flux ratios from the literature, including
recent \textit{Chandra} data, with new observations from the SMARTS
telescope and \textit{HST}, and use them to investigate the spatial
structure of the central regions using a Bayesian Monte Carlo analysis
of the microlensing variability. The wide wavelength coverage allows
us to constrain not only the accretion disk half-light radius
$r_{1/2}$, but the power-law slope $\xi$ of the size-wavelength
relation $r_{1/2} \propto \lambda^\xi$. With a logarithmic prior on
the source size, the (observed-frame) $R$-band half-light radius $\log
(r_{1/2}/\mathrm{cm})$ is $16.0\err{0.3}{0.4}$, and the slope $\xi$ is
$1.0\err{0.30}{0.56}$. We put upper limits on the source size in soft
(0.4$-$1.2\,keV) and hard (1.2$-$8\,keV) X-ray bands, finding 95\%
upper limits on $\log (r_{1/2}/\mathrm{cm})$ of 15.33 in both bands.
A linear prior yields somewhat larger sizes, particularly in the X-ray
bands. For comparison, the gravitational radius, using a black hole
mass estimated using the H$\beta$ line, is $\log(r_g/\mathrm{cm}) =
13.94$. We find that the accretion disk is probably close to face-on,
with $\cos i = 1.0$ being four times more likely than $\cos i =
0.5$. We also find probability distributions for the mean mass of the
stars in the foreground lensing galaxy, the direction of the
transverse peculiar velocity of the lens, and the position angle of
the projected accretion disk's major axis (if not face-on).

\end{abstract}


\section{Introduction}
\label{sec:intro}

The detailed structure of the innermost regions of active galactic
nuclei (AGNs), between 1 and $\sim$1000 gravitational radii from the
central black hole, has remained observationally elusive. The
gravitational radius of a black hole of mass $10^9 M_\odot$ is about
10 AU, so these scales cannot even remotely be resolved at
cosmological distances. The ultraviolet (UV), optical, and
near-infrared continuum is thought to come from a geometrically thin,
optically thick accretion disk \citep{Shakura:1973p337,
  Novikov:1973p343}. Although this model has been quite successful in
explaining the X-ray spectra of stellar-mass black hole binaries
\citep[e.g.,][]{McClintock:2011p114009}, the same has not been true of
the UV/optical spectra of AGNs, in part because of complications
arising from line emission \citep[e.g.,][]{Blaes:2001p560}. The X-ray
continuum is non-thermal and is thought to arise from the inverse
Compton scattering of disk photons by a corona of hot electrons
\citep[e.g.,][]{Reynolds:2003p389}. The spatial structure of the X-ray
corona is not known. In addition, in many AGNs the presence of the
iron K$\alpha$ emission line indicates that X-rays are being reflected
from the accretion disk \citep[e.g.,][]{Fabian:1989p729,
  Laor:1991p90}.

The gravitational microlensing of lensed quasars has proven to be an
effective tool for measuring the properties of quasar accretion disks,
and is starting to become useful for studying the X-ray corona as
well. The time-dependent microlensing magnification (or
demagnification) of one or more images of a lensed quasar is moderated
by the finite size of the source, which smooths the complicated
caustic pattern of microlensing magnifications as the quasar passes
over it. This allows us to use the microlensing magnifications to
estimate the source size, and such work has shown that in general the
accretion disks are larger than would be expected from either thin
disk modeling or total flux arguments \citep{Pooley:2007p19,
  Anguita:2008p327, Morgan:2010p1129, Hainline:2012p104,
  JimenezVicente:2012p106}. Since the effective temperature of the
disk depends on radius, its apparent size depends on wavelength,
leading to a chromatic dependence of the microlensing magnification,
with the same quasar image experiencing larger variability at blue
wavelengths than at red wavelengths. Several studies have used this to
constrain the power-law slope of the size-wavelength relation, and the
results have been consistent with each other and with the thin disk
prediction that the size goes like the 4/3 power of the wavelength,
mostly because of their large uncertainties \citep{Poindexter:2008p34,
  Bate:2008p1955, Eigenbrod:2008p933, Floyd:2009p233,
  Blackburne:2011p34, Mosquera:2011p145}. Finally, efforts to put
upper limits on the size of the X-ray regions have also been
successful \citep{Pooley:2006p67, Pooley:2007p19, Chartas:2009p174,
  Dai:2010p278}, and recently there have been attempts to constrain
the direct and reflected components' sizes separately using color cuts
or spectral decomposition \citep{Chen:2011pL34, Blackburne:2011p0027,
  Morgan:2012p52, Chen:2012p24, Chartas:2012p137, Mosquera:2013p5009}.

In this paper we use infrared (IR), optical, UV, and X-ray photometry
of the lensed $z_S = 2.32$ quasar
\heoneonelong\ \citep{Wisotzki:1993pL15, Wisotzki:1995pL59} to
quantitatively constrain the properties of its central emission
regions. \heoneone\ is lensed by a foreground early-type galaxy at
redshift $z_L = 0.73$ into a pair of images separated by
$3\farcs2$. The mass of the central black hole in \heoneone\ has been
investigated using the widths of the emission lines C\textsc{iv},
H$\beta$, and H$\alpha$, yielding mass estimates
$\log(M_\mathrm{BH}/M_\odot) = 9.37\pm0.33$, $8.77\pm0.30$, and
$9.05\pm0.23$, respectively \citep{Peng:2006p616, Greene:2010p937,
  Assef:2011p93}. We adopt the H$\beta$ mass of \citet{Assef:2011p93}
for this paper. We compare the light curves to microlensing
simulations using the Bayesian Monte Carlo method of
\citet{Kochanek:2004p58} and \citet{Poindexter:2010p668,
  Poindexter:2010p658}, which allows us to derive posterior
probability distributions for our parameters of interest, which
include the half-light radius of the accretion disk and the X-ray
emission regions, the slope with which the radius changes with
wavelength, and the inclination of the disk, as well as the mean mass
of the stars in the foreground galaxy.

\section{Multi-wavelength Data}
\label{sec:data}

We combine data from the literature with new photometry from the Small
and Moderate Aperture Research Telescope System (SMARTS) and
\textit{Hubble Space Telescope} (\textit{HST}) to create a lightcurve
spanning about 19 years and from the near-IR to X-rays in wavelength.

The previously-published data in our light curve come from
\citet[][$J$, $K$]{Courbin:1998p57}, 
\citet[][$V$, $I$, $K$]{Remy:1998p379}, 
\citet[][$H$]{Lehar:2000p584},
\citet[][$B$]{GilMerino:2002p428}, 
\citet{Schechter:2003p657},
\citet[][$V$]{Wyrzykowski:2003p229}, 
\citet[][$R$]{Ofek:2003p101},
\citet[][$H$, $K$, IRAC 3.6\,$\mu$m]{Poindexter:2007p146}, and
\citet[][F330W, F435W, $V$, F625W, $I$]{Munoz:2011p67}.
Where applicable, we have used the shorthand $V$, $I$, or $H$ for the
\textit{HST} filters F555W, F814W, or F160W, respectively. Although
\citet{Poindexter:2007p146} report flux ratios from the Spitzer Space
Telescope at several mid-IR wavelengths, we only use the 3.6\,$\mu$m
ratio, as it is the most likely to originate in the accretion disk
rather than a dusty torus. The mid-IR flux ratios are all nearly
identical, indicating that the source is large enough at all these
wavelengths for microlensing not to be important. We also note that we
have actually taken the \textit{HST} F160W magnitudes attributed to
\citet{Poindexter:2007p146} from the \textsc{Castles} database, since
due to a typographical error this paper reports the values of
\citet{Lehar:2000p584} rather than its own measurements. We have
confirmed that the \textsc{Castles} magnitudes correspond to the same
observations (E. Falco, private communication).

\subsection{Optical data}
\label{sec:optdata}

In addition, we use data from 8 seasons of monitoring by the ANDICAM
camera \citep{Depoy:2003p827} on the SMARTS telescope, primarily in
the $R$ and $J$ bands, with some data in $B$, $V$, and $I$. These
images are bias-corrected and flat-fielded by an automated pipeline,
and we stack the three to six images obtained on each night of
observation to improve the signal to noise ratio and reject cosmic
rays. We reject epochs with bad seeing (FWHM > 2\farcs{}0) or high sky
levels indicative of clouds or excessive moonlight. We measure the
fluxes of the quasar images using the point spread function (PSF)
fitting method described by \citet{Kochanek:2006p47}. The first 3
seasons of the $R$ and $J$ light curves are reported by
\citet{Poindexter:2007p146}, but for convenience we report them in
their entirety, together with the $B$, $V$, $I$, and $J$ data, in
Table~\ref{tab:smartslc}. Since we have re-analyzed the images, some
of the magnitudes differ slightly from the \citet{Poindexter:2007p146}
values.

\ifx \emulmacro \undefined
\else
\begin{deluxetable}{cccc}
\tablewidth{0pt}
\tablecaption{SMARTS Light Curves
  \label{tab:smartslc}}
\tablehead{
  \colhead{$\mathrm{HJD}-2450000$} &
  \colhead{A} &
  \colhead{B} &
  \colhead{Filter}
} 
\startdata
$2976.806$ & $+0.11\pm0.01$ & $+1.67\pm0.03$ & $J$  \\
$2976.807$ & $-0.01\pm0.01$ & $+1.38\pm0.02$ & $R$  \\
$2985.815$ & $+0.13\pm0.01$ & $+1.56\pm0.03$ & $J$  \\
$2985.818$ & $-0.02\pm0.01$ & $+1.39\pm0.01$ & $R$  \\
$2993.822$ & $+0.11\pm0.01$ & $+1.53\pm0.02$ & $J$  \\
$2993.826$ & $-0.03\pm0.02$ & $+1.40\pm0.02$ & $R$  \\
$3000.816$ & $+0.10\pm0.01$ & $+1.63\pm0.02$ & $J$  \\
$3000.820$ & $-0.05\pm0.02$ & $+1.40\pm0.02$ & $R$  \\
$3009.812$ & $+0.06\pm0.01$ & $+1.61\pm0.02$ & $J$  \\
$3009.815$ & $-0.09\pm0.01$ & $+1.41\pm0.02$ & $R$
\enddata
\tablecomments{~Light curves are in uncalibrated
magnitudes. This table is published in its entirety online. A portion
is shown here for guidance regarding its form and content.}
\end{deluxetable}

\fi

\heoneone\ has a fairly long lensing time delay between its two
images, which means that we must be wary of intrinsic quasar
variability conspiring with this delay to mimic microlensing
variability. For this work, we adopt a delay of 162.2 days
\citep{Morgan:2008p80}, with image A leading. For datasets (specified
uniquely by publication, observatory, and filter) with many
observations, we use linear interpolation on the light curve of image
B to remove the time delay. We do not extrapolate outside the bounds
of any dataset, and we limit the interpolation to dates within 30 days
of an actual data point, in order to avoid interpolating across
seasonal gaps. We also avoid comparing light curves from different
datasets, in order to avoid systematic calibration errors. Because of
this, we do not have to worry about the flux calibration of our data,
working exclusively with magnitude differences, or flux ratios between
the two quasar images. After interpolating, we bin sequential pairs of
observations if they are separated by 30 days or less. This is just to
keep our light curve short, since our simulation software's memory
requirements grow linearly with the number of epochs. Several
datasets, particularly the early ones, have single-epoch observations,
so we are forced to use the original (not time delay-corrected) flux
ratios, and to add to their uncertainties some estimate of the error
caused by intrinsic variability. For this estimate we use the
root-mean-square (RMS) of the difference between the delay-corrected
and uncorrected $R$-band light curves of image B. This value comes out to
0.11 mags, so for the affected observations we add 0.078 mags in
quadrature to the uncertainties for images A and B, dividing the total
between the two.

Several datasets, particularly the earlier ones, are too sparsely
sampled to allow linear interpolation according to our standard
procedure, but nevertheless have better coverage than the single-epoch
datasets. In these cases we choose pairs of epochs that are fairly
well-matched after correcting for the time delay, and we associate
image A from one epoch with image B from another, augmenting their
error bars as in the single-epoch cases, but by only half the
amount. In particular, we apply this method to a previously
unpublished pair of observations of \heoneone\ in the $B$ band from
the Instituto de Astrof\'isica de Canarias (IAC) 80-cm telescope at
the Observatorio del Teide. The respective instrumental magnitudes of
images A and B were $-1.179\pm0.016$ and $0.509\pm0.040$ on 19 January
1999, and $-1.261\pm0.036$ and $0.445\pm0.100$ on 18 March 1999. We
construct a single magnitude difference from these two observations,
adopting the earlier value for image B and the later value for image
A.

\ifx \emulmacro \undefined
\else
\ifx \emulmacro \undefined
\begin{deluxetable}{cccccc}
\else
\begin{deluxetable*}{cccccc}
\fi
\tablewidth{0pt}
\tablecaption{HST F275W Light Curves
  \label{tab:uvdata}}
\tablehead{
  \colhead{$\mathrm{HJD}-2450000$} &
  \colhead{A} &
  \colhead{B} &
  \colhead{Star a} &
  \colhead{Star b} &
  \colhead{Star c}
} 
\startdata
$5169.817$ & $17.529\pm0.002$ & $18.085\pm0.002$ & $16.094\pm0.001$ & $16.014\pm0.001$ & $18.170\pm0.002$ \\
$5227.008$ & $17.184\pm0.001$ & $18.123\pm0.002$ & $16.074\pm0.001$ & $16.004\pm0.001$ & $18.637\pm0.003$ \\
$5270.614$ & $17.113\pm0.001$ & $18.219\pm0.002$ & \nodata          & \nodata          & $18.831\pm0.003$ \\
$5340.061$ & $17.126\pm0.001$ & $18.589\pm0.003$ & $16.113\pm0.001$ & $16.043\pm0.001$ & $18.344\pm0.002$ \\
$5382.058$ & $16.904\pm0.001$ & $18.327\pm0.002$ & $16.086\pm0.001$ & $16.008\pm0.001$ & $18.223\pm0.002$ \\
$5505.015$ & $17.348\pm0.001$ & $18.328\pm0.002$ & $16.094\pm0.001$ & $16.012\pm0.001$ & $18.345\pm0.002$ \\
$5546.026$ & $17.412\pm0.001$ & $18.298\pm0.002$ & $16.088\pm0.001$ & $16.010\pm0.001$ & $18.267\pm0.002$ \\
$5654.585$ & $17.242\pm0.001$ & $18.023\pm0.002$ & $16.107\pm0.001$ & $16.026\pm0.001$ & $18.397\pm0.002$ \\
$5729.072$ & $17.207\pm0.001$ & $18.258\pm0.002$ & $16.113\pm0.001$ & $16.048\pm0.001$ & $18.645\pm0.003$ \\
$5877.216$ & $17.147\pm0.001$ & $18.047\pm0.002$ & $16.090\pm0.001$ & $16.013\pm0.001$ & $18.219\pm0.002$
\enddata
\tablecomments{~Light curves are in ST magnitudes. In this filter, the
offset from ST to AB magnitudes is $m_\mathrm{AB}-m_\mathrm{ST} =
1.532$.}
\ifx \emulmacro \undefined
\end{deluxetable}
\else
\end{deluxetable*}
\fi

\fi

We give the $V$-band data of \citet{Schechter:2003p657} and
\citet{Wyrzykowski:2003p229} special treatment in two ways. First, due
to its dense sampling we bin it four observations at a time rather
than pairwise, and second, we add in quadrature an extra uncertainty
of 0.03 mags to each image. This is about the level of the
high-frequency variability seen in image A in these data, and
interpreted variously as microlensing of relativistic knots
\citep{Schechter:2003p657}, echoes of the intrinsic quasar variability
due to luminosity-dependent accretion disk area
\citep{Blackburne:2010p1079}, or microlensing of stochastic hot spots
on the disk \citep[e.g.,][]{Dexter:2011pL24}. We do not currently have
the capability to test these hypotheses (at least, not while running a
multiwavelength simulation), so we add the extra uncertainty to avoid
skewing our results by trying to fit this variability with stellar
microlensing.

\ifx \emulmacro \undefined
\else
\begin{figure*}
  \includegraphics[width=\textwidth]{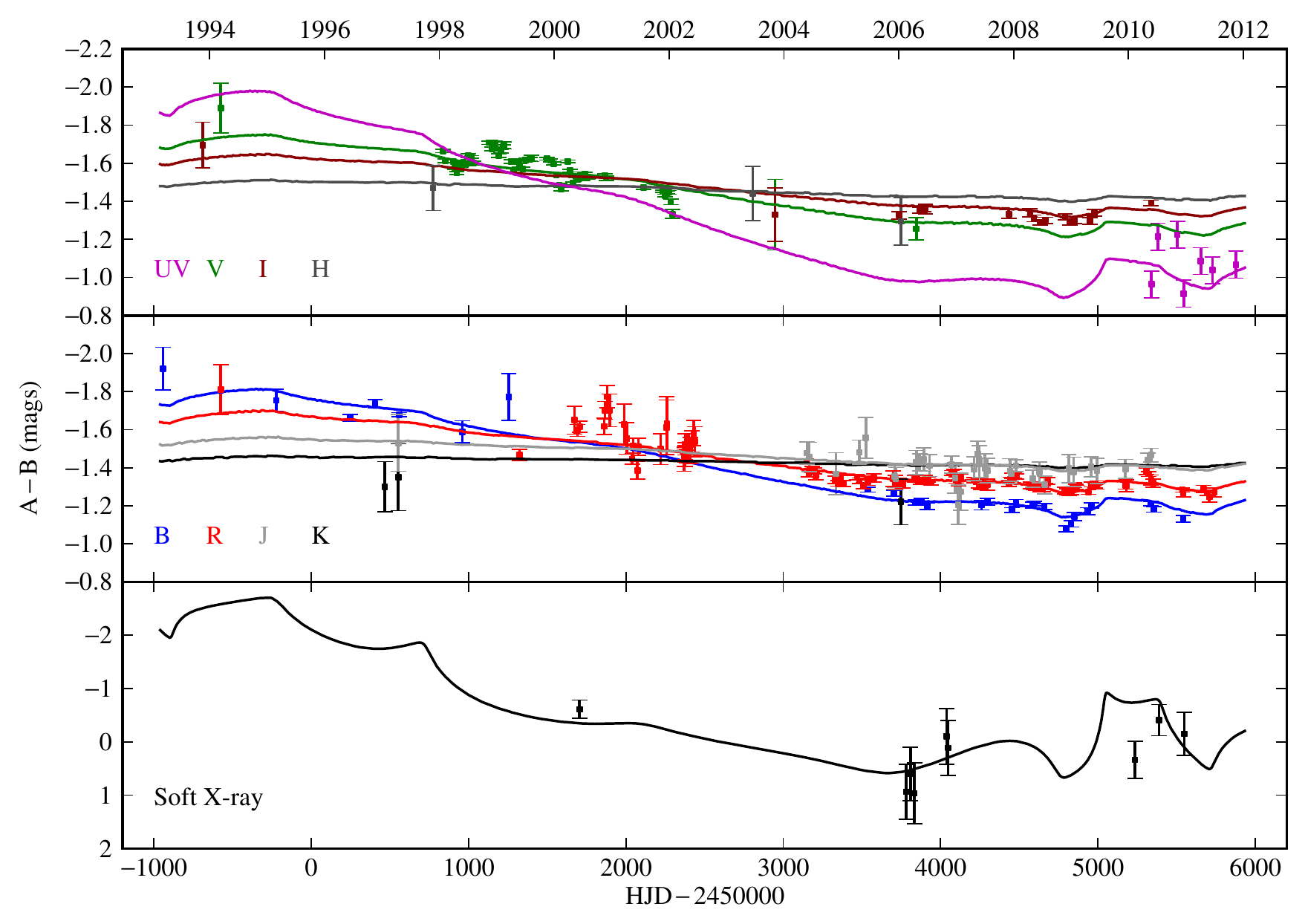}
  \caption{Time delay-corrected difference in magnitudes between the A
    and B images of \heoneone\ in each filter as a function of time
    (points with error bars). Some systematic errors have been added
    as described in Section~\protect{\ref{sec:data}}. Curves show one
    of the best-fitting out of millions of microlensing light curves
    found by our simulation. The soft X-ray light curve is in its own
    panel because of its much larger range. The hard X-ray data are
    nearly identical, and we do not show them.}
  \label{fig:lc}
\end{figure*}
\fi

Our $B$-band data includes flux from the redshifted Ly$\alpha$
emission line, and $R$, $H$, and $K$ are also somewhat affected by the
C\textsc{iii]}, H$\beta$, and H$\alpha$ lines, respectively. But in
these broad filters, the line flux only amounts to a few percent of
the total flux, and we do not expect it to have a strong effect on the
microlensing results \citep{Dai:2010p278}.

The light curve of the magnitude difference of the two quasar images,
with error bars as described in this section, is shown for each band
in Figure~\ref{fig:lc}, together with an example of one of the many
microlensing light curves that we fit to the data. Since this is the
flux ratio of the two quasar images, and is corrected for the time
delay, the variability is due to microlensing. The trend of increasing
microlensing variability with decreasing wavelength, which is
qualitatively expected for an accretion disk, can be easily seen.

\subsection{UV Data}
\label{sec:uvdata}

We have observed \heoneone\ in the F275W filter (rest-frame
$0.083$\,$\mu$m) using the UVIS channel of the Wide Field Camera 3
aboard \textit{HST} at 10 epochs roughly evenly spaced between 2009
December 4 and 2011 November 11. At each epoch, we combine four images
totalling 2524 seconds of exposure time using the Multidrizzle task
within the PyRAF software package\footnote{PyRAF and Multidrizzle are
  products of the Space Telescope Science Institute, which is operated
  by AURA for NASA.}. We take care to prevent the automatic cosmic ray
rejection from masking out the PSF cores of the quasar components or
bright comparison stars, manually removing some pixels from the bad
pixel mask before repeating the final drizzle. The resulting images
are clean and contain little besides the two quasar images and a
handful of other pointlike objects.

We use aperture photometry to determine the fluxes of the quasar
images and three other comparison objects. This is a very simple
process because of the paucity of sources and relatively large
separation between the quasar images. We use a square $1\farcs{}6$
aperture, and measure the sky background in a square annulus just
outside the aperture, with an outer diameter twice that of the
aperture. We then use the header keywords PHOTFLAM and PHOTZPT to
convert the instrumental magnitudes to the ST system (in this filter,
the offset from ST to AB magnitudes is $m_\mathrm{AB} - m_\mathrm{ST}
= 1.532$). Table \ref{tab:uvdata} lists the ST magnitudes and formal
uncertainties of quasar images A and B, as well as three comparison
objects that we label a, b, and c. The first two of these, a and b,
are the bright stars northeast of the lens labeled 4 and 3
(respectively) by \citet{Wisotzki:1995pL59}, and the last is located
at a position $(\Delta \alpha \cos \delta, \Delta \delta) = (-55\farcs{}7,
-38\farcs{}2)$ relative to quasar image A. Objects a and b show little
variability, while object c varies by $\sim$$0.3$\,mags and may well
be a quasar.

Since \heoneone\ has a relatively high redshift, these UV observations
probe the far-UV region of the quasar spectrum between the Lyman limit
and the Ly$\beta$ emission line, from $\lambda \simeq 76$ to
91\,\AA\ in the rest frame. Though there may be some contamination
from the remainder of the Lyman series, this is unlikely to be
strong, and we consider the majority of the UV flux to be coming from
the quasar accretion disk.

We shift the light curve of image B and use linear interpolation to
estimate its delay-corrected light curve, resulting in a 7-epoch light
curve. We add 0.05 mags in quadrature to the uncertainties of each
image to account for the systematics introduced by this
process. Fortunately, the mismatch between each measurement of image A
and the nearest shifted measurement of B is usually fairly small, on
the order of 10-30 days. The UV flux ratios are shown in the top panel
of Figure~\ref{fig:lc}.

\subsection{X-Ray data}
\label{sec:xraydata}

For our investigation of the X-ray properties of the quasar, we use
the X-ray light curves of \heoneone\ from \citet{Chen:2012p24} and
\citet{Chartas:2009p174}.  These data consist of soft-band
(0.4$-$1.2\,keV) and hard-band (1.2$-$8\,keV) absorption-corrected
count rates at nine epochs. We convert the count rates to instrumental
magnitudes and symmetrize the resulting error bars by taking the
geometric mean of the upper and lower errors.

The sampling of the X-ray light curves is generally not well-matched
to the time delay between the two quasar images. In particular, the
first epoch is separated from the rest by a period much longer than
the delay. For this epoch we do not apply any delay correction, but
simply augment the error bars in quadrature by 0.078 mags as
previously described. For the remaining epochs we linearly interpolate
the light curve of image B, then add the same extra uncertainty,
except for the last two epochs. For these epochs, the observational
cadence is within 10 days of the time delay, so that we can easily
match one observation of B with the next observation of A. We add only
half the extra uncertainty (0.039 mags) in these cases. The resulting
soft-band X-ray flux ratios are shown in the bottom panel of
Figure~\ref{fig:lc}.

Unlike some other lensed quasars such as \qtwotwolong\
\citep{Chen:2011pL34} and \rxoneonelong\ \citep{Chartas:2012p137},
\heoneone\ does not show significant differences between its hard and
soft X-ray light curves. This immediately indicates that our
microlensing simulation analysis will not be able to give definitive
evidence of a difference in source size for these two bands. However,
we ought to be able to place upper limits on the magnitude of the
logarithm of their size ratio, which is an interesting result in its
own right.

\section{Microlensing Simulations}
\label{sec:simulations}

We model the gravitational microlensing of the two quasar images using
the Bayesian Monte Carlo technique described by
\citet{Kochanek:2004p58} and updated by \citet{Poindexter:2010p668,
  Poindexter:2010p658}. This method uses ray-tracing to generate
magnification patterns that encode the microlensing magnification
experienced by each quasar image on top of its magnification from the
``macro''-lensing by the lens galaxy as a whole. We generate new
patterns for each epoch in the quasar's light curve, allowing the
stars to move between epochs. This causes the pattern of
high-magnification caustics and low-magnification troughs to evolve.
Together with the bulk relative motion of the quasar and the lens
galaxy, this leads to variability in the microlensing
magnifications. This work is the first to simultaneously fit
observations at a variety of wavelengths and include the effects of
stellar motions.

To generate the magnification patterns, we scatter stars randomly
across a portion of the image plane. The mass function that we use for
the microlens stars is a power law with a slope $d\ln N/d\ln M = -1.3$
and a mass range $M_\mathrm{max}/M_\mathrm{min} = 50$. We vary the
mean mass $\langle M \rangle$ between $0.03 M_\odot$ and $10 M_\odot$
in six logarithmically spaced steps. At the location of each quasar
image, the fraction of the surface mass density made up of stars (as
opposed to smoothly distributed dark matter) is determined from a
global parameter, $f_{M/L}$. This parameter varies between 0 (no
stars) and 1 (all stars), and specifies the amplitude of the stellar
component (in a lensing model consisting of a de Vaucouleurs stellar
component and a Navarro-Frenk-White dark matter component) relative to the
best-fitting model with only stars. These models come from the work of
\citet{Poindexter:2007p146}, and are constrained not only by the
relative positions of the quasar images and lens galaxy, but by the
images' near-IR fluxes, which are not strongly affected by
microlensing, and by their estimate of the lensing time delay. They
find that a value of $f_{M/L} = 0.3$ is most likely; this estimate
would probably not change much if more recent time delay estimates
were used instead. In the interest of fair sampling, we allow
$f_{M/L}$ to take the values 0.1, 0.3, and 1.0. From these same lens
models we obtain the total lensing convergence $\kappa_\mathrm{tot}$
and shear $\gamma$ at the positions of the quasar images, which
together with $\langle M \rangle$ and $f_{M/L}$ determine the overall
character of the magnification patterns. We set the rest-frame
one-dimensional velocity dispersion of the stars to 301\,km\,s$^{-1}$;
this is calculated from the monopole strength of a singular isothermal
sphere plus external shear (SIS$\gamma$) model for the lens
potential. This has been shown to be a good estimator for the stellar
velocity dispersion \citep{Treu:2006p662}.

The effect of the quasar's finite size is simulated by convolving the
magnification patterns with a source light profile. We use a profile
with surface brightness $I(r) \propto [\exp[(r/r_\lambda)^{3/4}] -
  1]^{-1}$, where $r_\lambda$ is defined by the relation $k
T_\mathrm{eff}(r_\lambda) = h_p c / \lambda$. In standard thin disk
theory,
\begin{align}
r_\lambda &= \left(\frac{45 G \lambda^4 M_\mathrm{BH}\dot{M}}
{16 \pi^6 h_p c^2}\right)^{1/3} \nonumber \\
&= \left( 9.7 \times 10^{15}\,\mathrm{cm} \right)
\left( \frac{\lambda}{\mu\mathrm{m}} \right)^{4/3}
\left( \frac{M_\mathrm{BH}}{10^9M_\odot} \right)^{2/3}
\left( \frac{L}{\eta L_\mathrm{Edd}} \right)^{1/3} ~,
\end{align}
 where $\lambda$ is the rest wavelength, $L_\mathrm{Edd}$ is the
 Eddington luminosity, and $\eta$ is the radiative efficiency. In this
 model, the disk is a multicolor blackbody with temperature
 $T_\mathrm{eff} \propto r^{-\beta}$ and $\beta = 3/4$. We neglect the
 effects of the inner edge of the disk. We vary the projected area of
 the disk at a fixed wavelength corresponding to the observed-frame
 $R$ band, allowing it to take values separated by 0.2 dex. We present
 most of our results, however, in terms of the half-light radius of
 the disk $r_{1/2}$, which is proportional to the square root of the
 area divided by the cosine of the inclination angle, and is equal to
 $2.44 r_\lambda$ when $\beta = 3/4$. The half-light radius is assumed
 to vary as a power law with wavelength, $r_{1/2} \propto
 \lambda^\xi$, and we allow its power-law slope $\xi$ to vary as
 well. Generically, $\xi$ is the reciprocal of the temperature slope,
 so for a thin disk its value is $4/3$. We allow it to take values
 ranging from $-1$ to $+2$ in 7 steps. Allowing the slope to vary is
 inconsistent with our choice of $\beta=3/4$ for the light profile of
 the source, but fortunately the details of the light profile, apart
 from the half-light radius, do not have a strong effect on the
 microlensing \citep{Mortonson:2005p594}. Since the X-ray flux does
 not arise from the same blackbody source as the UV/optical flux, we
 do not assume that the X-ray size lies on the same power-law curve,
 instead allowing the area of the source in soft and hard X-rays to
 vary independently. We do this by pursuing an iterative strategy,
 first performing an initial simulation that is ignorant of the X-ray
 data, and then following up with secondary simulations for soft and
 hard X-ray data, each of which makes use of the output of the
 preceding simulation. The secondary simulations do not generate
 random trajectories for the quasar across the magnification patterns;
 instead they use the trajectories that successfully fit the data in
 the previous simulation, and they only vary a single new parameter:
 the area of the X-ray source. We do not take into account the lensing
 of the X-ray flux by the black hole, though given the compactness of
 the X-ray source it may be advisable to include this effect in future
 work \citep{Chen:2012p6487}.

We vary the inclination of the accretion disk between $\cos i = 1$
(face-on) and $\cos i = 0.2$ (nearly edge-on) in 5 steps evenly
distributed in $\cos i$; this is equivalent to a uniform prior on the
orientation of the disk in three dimensions. Since an inclined disk
looks like an ellipse to the observer, we also let the position angle
of the projected disk take 9 values evenly distributed between $\phi_a
= 0^\circ$ (major axis aligned with north) and $\phi_a = 160^\circ$
(rotated east of north). These parameters have subtle effects on the
smoothed magnification patterns, and \citet{Poindexter:2010p668} show
that with moving patterns it is possible to constrain their
values. But we note that they are helped by the relative importance of
the random stellar motions in the lens system they consider
(\qtwotwo), and the only other attempt to use this method so far does
not obtain strong constraints on these parameters for the lens
\hefourlong\ \citep{Blackburne:2011p0027}.

As described by \citet{Kochanek:2004p58} and
\citet{Poindexter:2010p668, Poindexter:2010p658}, this analysis method
generates billions of trial paths that the quasar may take across the
magnification patterns, and evaluates the likelihood of each trial
based on the goodness of fit of the resulting light curve. The
calculations are described in Equations~(4) and (5) of
\citet{Blackburne:2011p0027}; we set the ``rescale'' parameter $f_0^2$
to 2.0, 1.0, and 1.0 for the UV/optical, soft X-ray, and hard X-ray
bands, respectively. Since each trial is associated with some location
in parameter space, this Monte Carlo sampling gives us a reasonable
idea of the joint likelihood function of our parameters. We then
multiply the likelihood by our priors to obtain a Bayesian posterior
probability density functions (PDFs) for our parameters. Our priors
are mostly uniform, either in linear or logarithmic space. The mean
microlens mass $\langle M \rangle$ and the variable parameterizing the
fraction of the surface density in stars $f_{M/L}$ have logarithmic
priors, and the inclination $\cos i$, the disk position angle
$\phi_a$, and the wavelength slope $\xi = d\ln r_{1/2}/d\ln \lambda$
have linear priors. For the half-light radius $r_{1/2}$ itself we use
both linear and logarithmic priors, because they can both be sensibly
applied \citep[see discussion by][]{Blackburne:2011p0027}, and in order
to evaluate the robustness of our results to changes in this
prior. For the starting positions of the quasar on the magnification
patterns we choose a simple uniform prior, and for its velocity we
choose a circular Gaussian. The velocity prior is calculated in the
manner described by \citet{Blackburne:2011p0027}, centering the
Gaussian on the projection onto the lens plane of the velocity of the
solar system relative to the cosmic microwave background (CMB), and
setting its width to the quadrature sum of estimates of the peculiar
velocities of the lens galaxy and the source galaxy. All velocities
are corrected for cosmological time-dilation, and are converted to
angular velocities using the angular diameter distances
$D_\mathrm{OL}$, $D_\mathrm{OS}$, and $D_\mathrm{LS}$ (observer to
lens, observer to source, and lens to source, respectively). Projected
into the lens plane, the effective one-dimensional width of the
velocity prior is
\begin{equation}
  \sigma_\mathrm{eff}^2 = \sigma_L^2 + \sigma_S^2
  \left(\frac{1+z_L}{1+z_S}\right)^2 
  \left(\frac{D_\mathrm{OL}}{D_\mathrm{OS}}\right)^2 ~.
\end{equation}
We use the prescription cited by \citet{Mosquera:2011p96} for the
estimates of the peculiar velocity dispersion, and find
$\sigma_L=275$\,km\,s$^{-1}$ and $\sigma_S=201$\,km\,s$^{-1}$,
yielding a width for the velocity prior of
$\sigma_\mathrm{eff}=290$\,km\,s$^{-1}$. This value is also
approximately the typical speed of the quasar across the pattern.

\section{Results}
\label{sec:results}

\ifx \emulmacro \undefined
\else
\begin{figure}
  \includegraphics[width=\columnwidth]{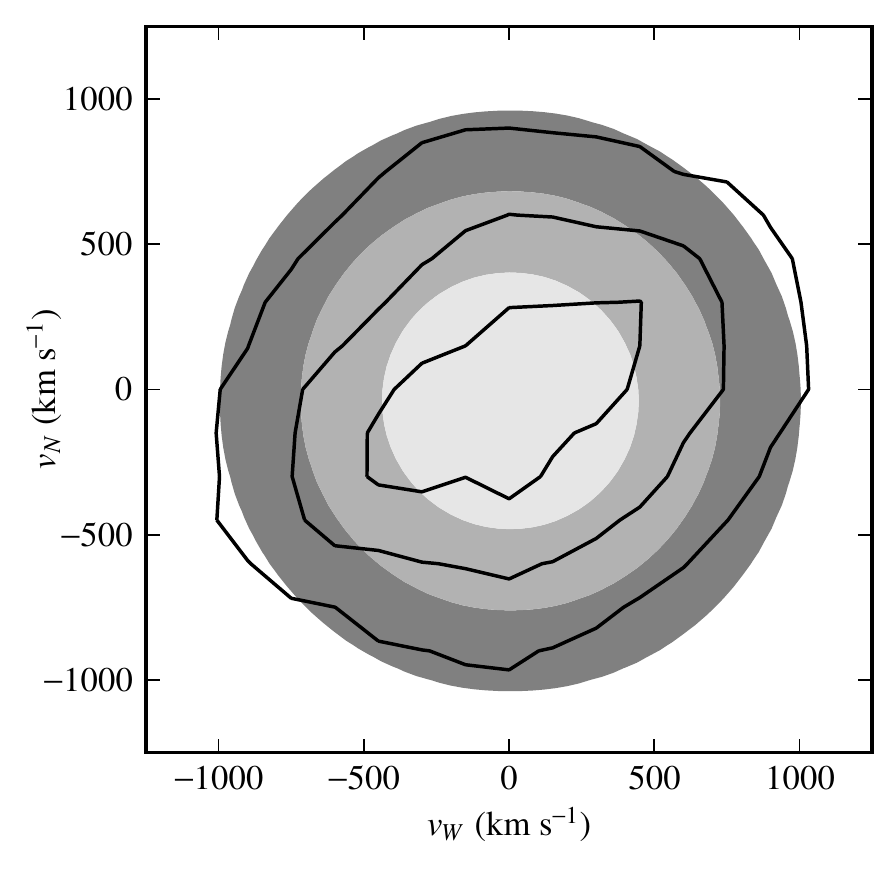}
  \caption{Filled contours indicate the 68\%, 95\%, and 99\% confidence
    levels of the velocity prior described in
    Section~\ref{sec:simulations}. Open contours show the posterior
    probability distribution of the transverse lens velocity (relative
    to that of the source and the observer). The direction of motion
    of the lens galaxy is $120\pm40$ degrees, with a 180-degree
    degeneracy due to the symmetry of the elliptical disk model.}
  \label{fig:vel}
\end{figure}
\fi

The product of the likelihood distribution with our priors gives a
joint posterior probability distribution for our parameters. In this
section, we examine the projections of this distribution, specifically
the posterior PDFs for the velocity of the quasar relative to the lens
galaxy, the inclination of the accretion disk and the position angle
of its major axis, the mean mass of the stars causing the
microlensing, the half-light radius of the quasar in the optical, soft
X-ray, and hard X-ray bands, and the power-law slope of the half-light
radius of the accretion disk with wavelength.

\subsection{Velocity}
\label{sec:velresult}

The moving magnification patterns break the strict degeneracy between
the velocity of the quasar relative to the lens and the mean mass of
the stars, and allows us to put some actual constraints on the
velocity. The posterior PDF for the transverse velocity of the lens
(relative to the observer and source) is shown, together with our
prior on the same quantity, in Figure~\ref{fig:vel}. More precisely,
this quantity is the angular velocity of the source across the
magnification pattern, multiplied by the lens distance $D_\mathrm{OL}$
and time-dilated to the lens redshift, with a change of sign to make
it a lens velocity rather than a source velocity. This differs
somewhat in magnitude from the transverse peculiar velocity of the
lens galaxy because it includes terms from the source and observer
motions, but the lens term dominates and the difference is therefore
small. Though the radial profile of the probability distribution is
mostly determined by the prior, we do see a preference for a direction
of motion of roughly $120 \pm 40$ degrees (68\% confidence), with a
180-degree degeneracy related to the bilateral symmetry of the
(generally inclined) accretion disk.

\subsection{Disk Inclination}
\label{sec:cosi}

\ifx \emulmacro \undefined
\else
\begin{figure}
  \includegraphics[width=\columnwidth]{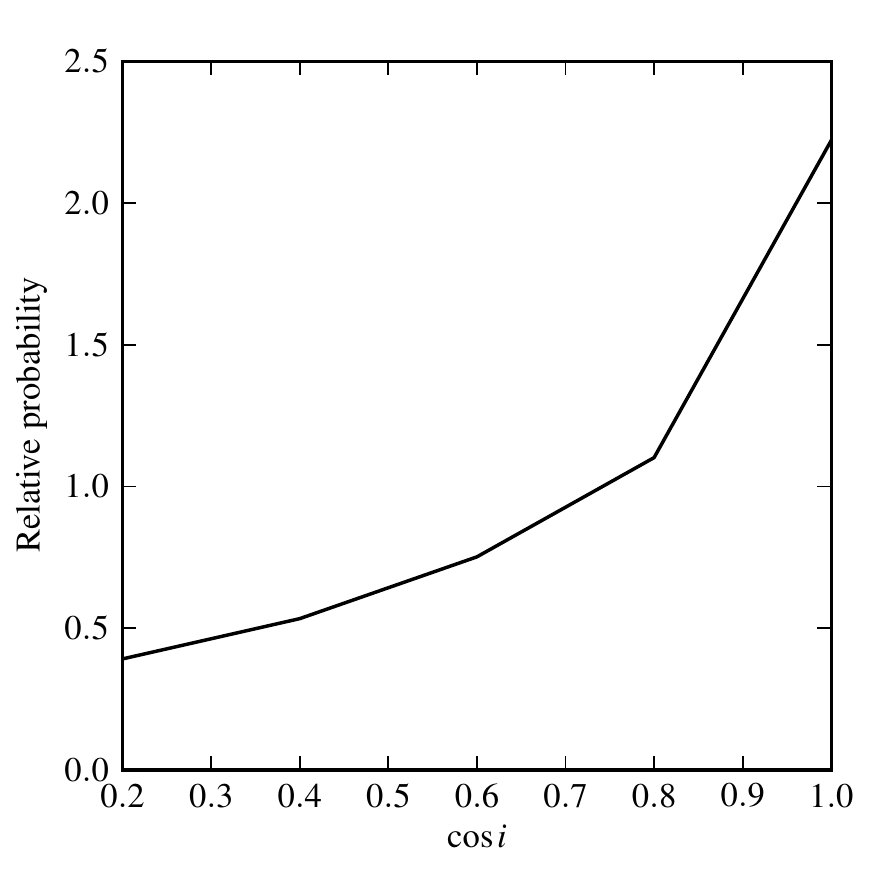}
  \caption{Posterior distribution for the accretion disk inclination
    $\cos i$. A face-on disk has $\cos i = 1$.}
  \label{fig:cosi}
\end{figure}
\fi

\ifx \emulmacro \undefined
\else
\begin{figure}
  \includegraphics[width=\columnwidth]{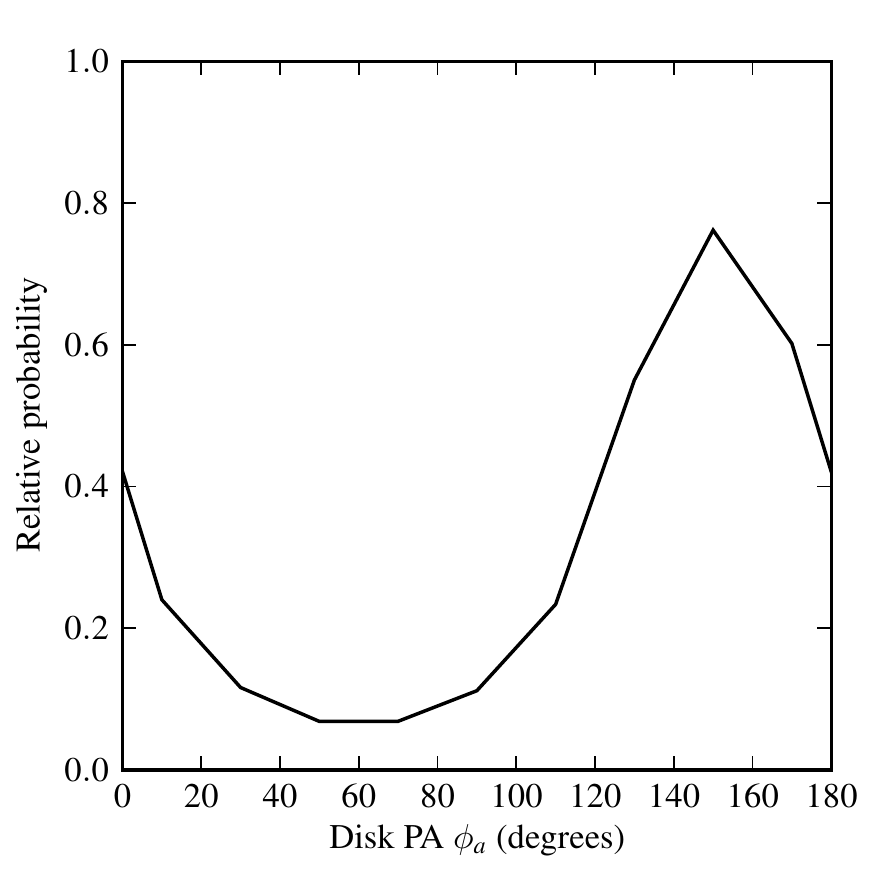}
  \caption{Posterior distribution for the accretion disk major axis
    position angle $\phi_a$, measured in degrees East of
    North. Face-on ($\cos i = 1$) solutions are excluded in
    calculating this result.}
  \label{fig:diskpa}
\end{figure}
\fi

\ifx \emulmacro \undefined
\else
\begin{figure}
  \includegraphics[width=\columnwidth]{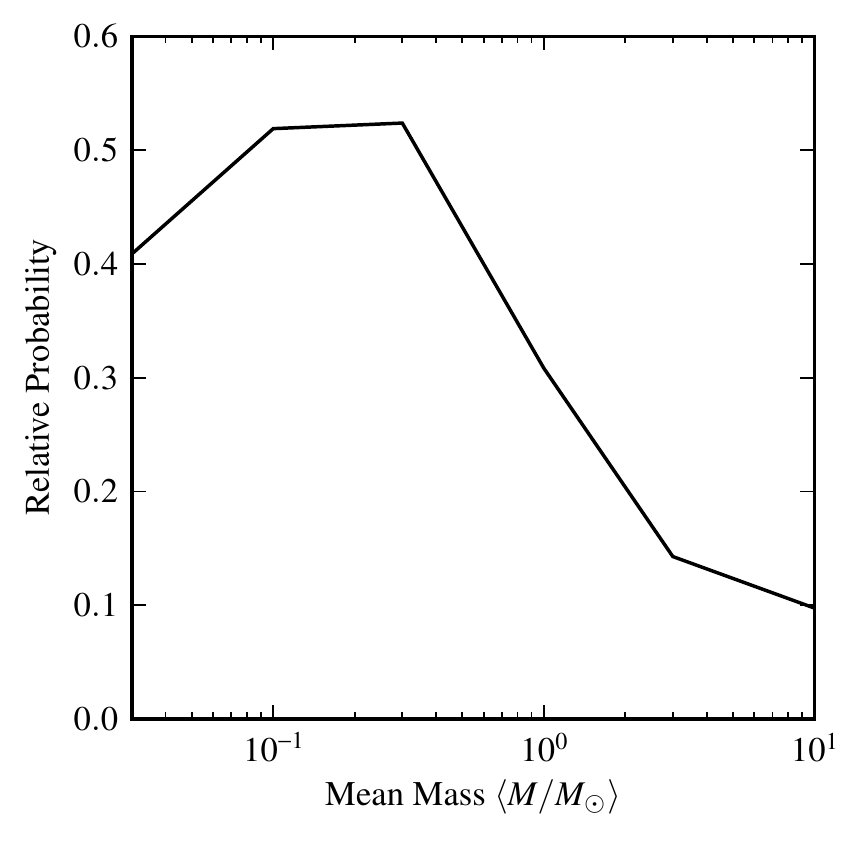}
  \caption{Posterior distribution for the mean mass of the stars in
    the lensing galaxy.}
  \label{fig:meanmass}
\end{figure}
\fi

The posterior PDFs for the accretion disk inclination $\cos i$ and the
position angle of its major axis $\phi_a$ are shown in
Figures~\ref{fig:cosi} and \ref{fig:diskpa}. A face-on disk is
favored, with $\cos i = 1.0$ about four times as likely as $\cos i =
0.5$. If we assume that the disk is in fact inclined and discard
trials with $\cos i=1.0$, then its major-axis position angle is fairly
well-determined, with a value of $150 \pm 33$ degrees East of North
(68\% confidence).

\subsection{Mean Microlens Mass}
\label{sec:meanmass}

Figure~\ref{fig:meanmass} shows the posterior PDF for the mean mass
$\langle M \rangle$ of the stellar microlenses. This quantity
determines the scaling between the natural angular units of the
magnification patterns (microlens Einstein radii) and units of
physical length (e.g., cm) in the source plane. Even with our moving
magnification patterns, it is still somewhat dependent on the velocity
prior, in the sense that the very large transverse velocities
suppressed by the Gaussian wings of our velocity prior correspond to
very large mean masses. But we are confident in the appropriateness of
this prior; in any case it seems quite unlikely that the mass function
of stars in the lensing galaxy has a mean higher than
1\,$M_\odot$. The distribution is fairly broad, and favors mean masses
of 0.1 to 0.3\,$M_\odot$.

\subsection{Accretion Disk Size}
\label{sec:sizes}

The posterior PDFs for the deprojected half-light radius of the
accretion disk in the $R$ band, and of the soft (0.4 to 1.2\,keV) and
hard (1.2 to 8\,keV) X-ray emission regions are shown in
Figure~\ref{fig:rhalf}. The logarithm of the $R$-band radius (measured
in cm) is $16.0\err{0.3}{0.4}$ with the logarithmic prior. With the
linear prior, this value rises to $16.2\pm0.3$. The X-ray PDFs do not
converge at small sizes, as the finite resolution of the magnification
patterns prevents us from going to smaller sizes, but the cutoff scale
is smaller than the gravitational radius of the black hole (regardless
of which black hole mass estimate is chosen). The three vertical lines
in Figure~\ref{fig:rhalf} show the gravitational radius, the $R$-band
radius estimated from the flux of the quasar, and the $R$-band radius
predicted by the thin disk model, assuming an accretion efficiency
$\eta=0.1$ and an Eddington fraction $L / L_\mathrm{Edd} = 1$. All
assume the H$\beta$ mass estimate of \citet{Assef:2011p93}. With the
logarithmic prior the source is smaller than $\log
(r_{1/2}/\mathrm{cm}) = 15.33$ at 95\% confidence in both soft and
hard X-rays. With the linear prior, the upper limits are 15.71 (soft)
and 15.59 (hard). If the \citet{Peng:2006p616} black hole mass
estimate is correct, then half the X-ray light is originating within a
radius less than about 6 (logarithmic prior) or 12-14 (linear prior)
gravitational radii.

\ifx \emulmacro \undefined
\else
\begin{figure}
  \includegraphics[width=\columnwidth]{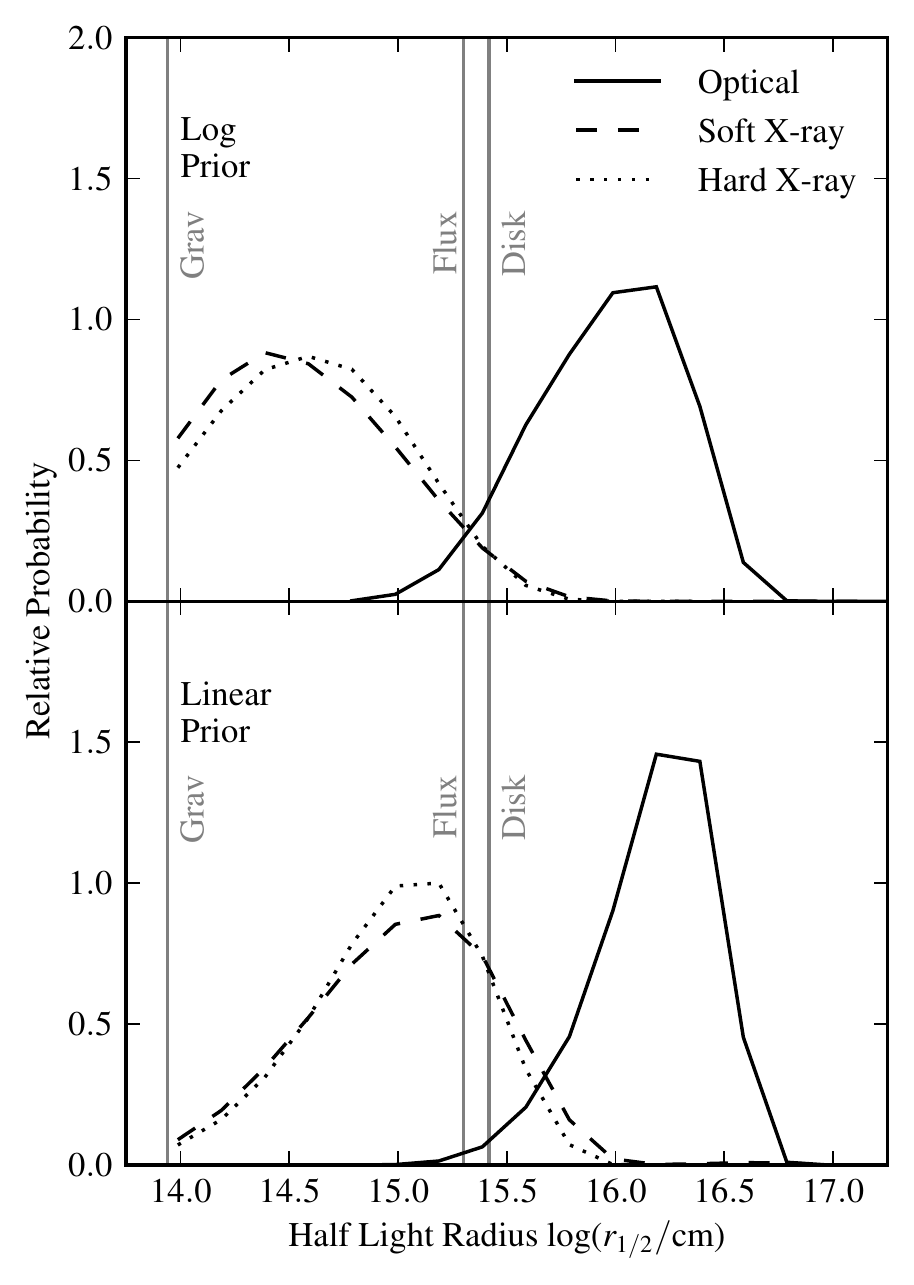}
  \caption{Posterior distribution for the half-light radius of the
    quasar accretion disk in the $R$ band and the two X-ray bands. In
    the upper panel we use a logarithmic prior, and in the lower panel
    we use a linear prior. The stellar mean mass $\langle M/M_\odot
    \rangle$ is fixed at 0.3. The gravitational radius of the black
    hole and the $R$-band size estimates based on the quasar flux and
    the thin disk model are shown as a vertical lines, marked
    ``Grav,'' ``Flux,'' and ``Disk'' respectively. The ``Grav'' and
    ``Disk'' values assume the H$\beta$ black hole mass estimate of
    \protect{\citet{Assef:2011p93}}, and can shift up to
    $\sim$0.5\,dex based on the uncertainties or by using the
    H$\alpha$ or \textsc{Civ} estimates.}
  \label{fig:rhalf}
\end{figure}
\fi

\ifx \emulmacro \undefined
\else
\begin{figure}
  \includegraphics[width=\columnwidth]{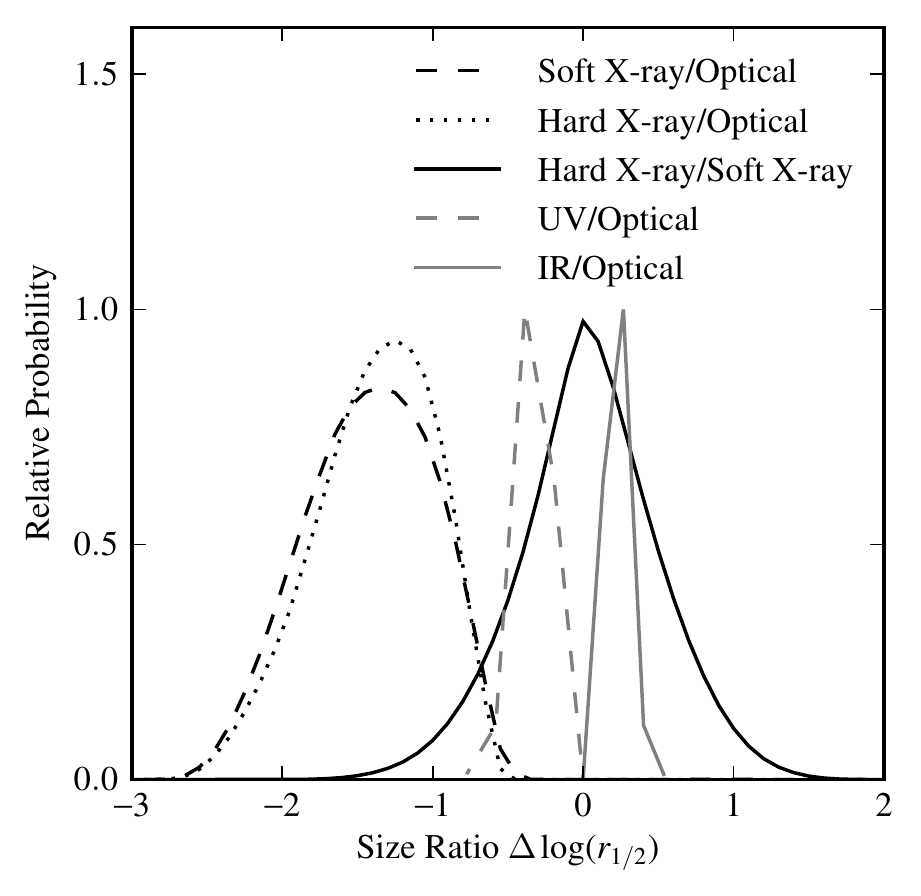}
  \caption{Posterior distributions for the ratios of the quasar's
    half-light radius. The black curves indicate ratios of the X-ray
    sizes; for these curves the mean mass $\langle M \rangle$ is set
    to $0.3 M_\odot$. The X-ray emission is much more compact than the
    observed-frame $R$-band emission (labeled ``Optical''). Also, our
    X-ray data rule out a very large difference in the sizes of the
    hard and soft X-ray sources. The gray curves indicate the size of
    the accretion disk at the (observed-frame) UV and $J$-band
    wavelengths, relative to the $R$-band size. Since we parameterize
    the disk using the $R$-band size and the wavelength slope $\xi$,
    the gray curves are simply scaled versions of the posterior
    distribution for $\xi$ (see \protect{Figure~\ref{fig:srcexp}}).}
  \label{fig:sizerat}
\end{figure}
\fi

\ifx \emulmacro \undefined
\else
\begin{figure}
  \includegraphics[width=\columnwidth]{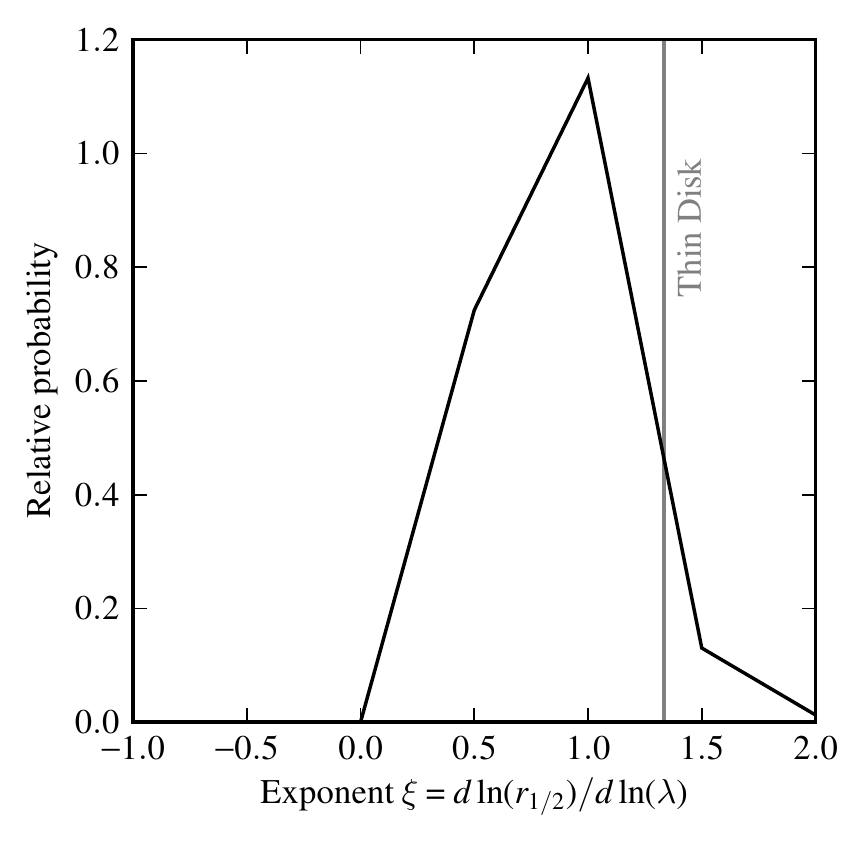}
  \caption{Posterior probability distribution for the power-law slope
    of the wavelength dependence of the half-light radius. The
    standard thin disk model predicts a value of 4/3 since
    $T_\mathrm{eff} \propto r^{-3/4}$.}
  \label{fig:srcexp}
\end{figure}
\fi

The deprojected half-light radii are calculated from the projected
area of the source $A = \pi (r_{1/2}/2.44)^2 \cos i$. With the
logarithmic (linear) prior, we find that $\log (A/\mathrm{cm}^2) =
31.54\err{0.64}{0.72}$ ($32.26\err{0.39}{0.50}$). Likewise, the 95\%
confidence upper limit on the logarithm of the source area is 30.23
(31.34) for soft X-rays, and 30.20 (31.11) for hard X-rays. These
projected areas are more appropriate for comparison with previous work
that does not vary the inclination of the source, but since \heoneone\
seems to have a relatively face-on disk, this correction is not very
important.

\ifx \emulmacro \undefined
\else
\begin{figure}
  \includegraphics[width=\columnwidth]{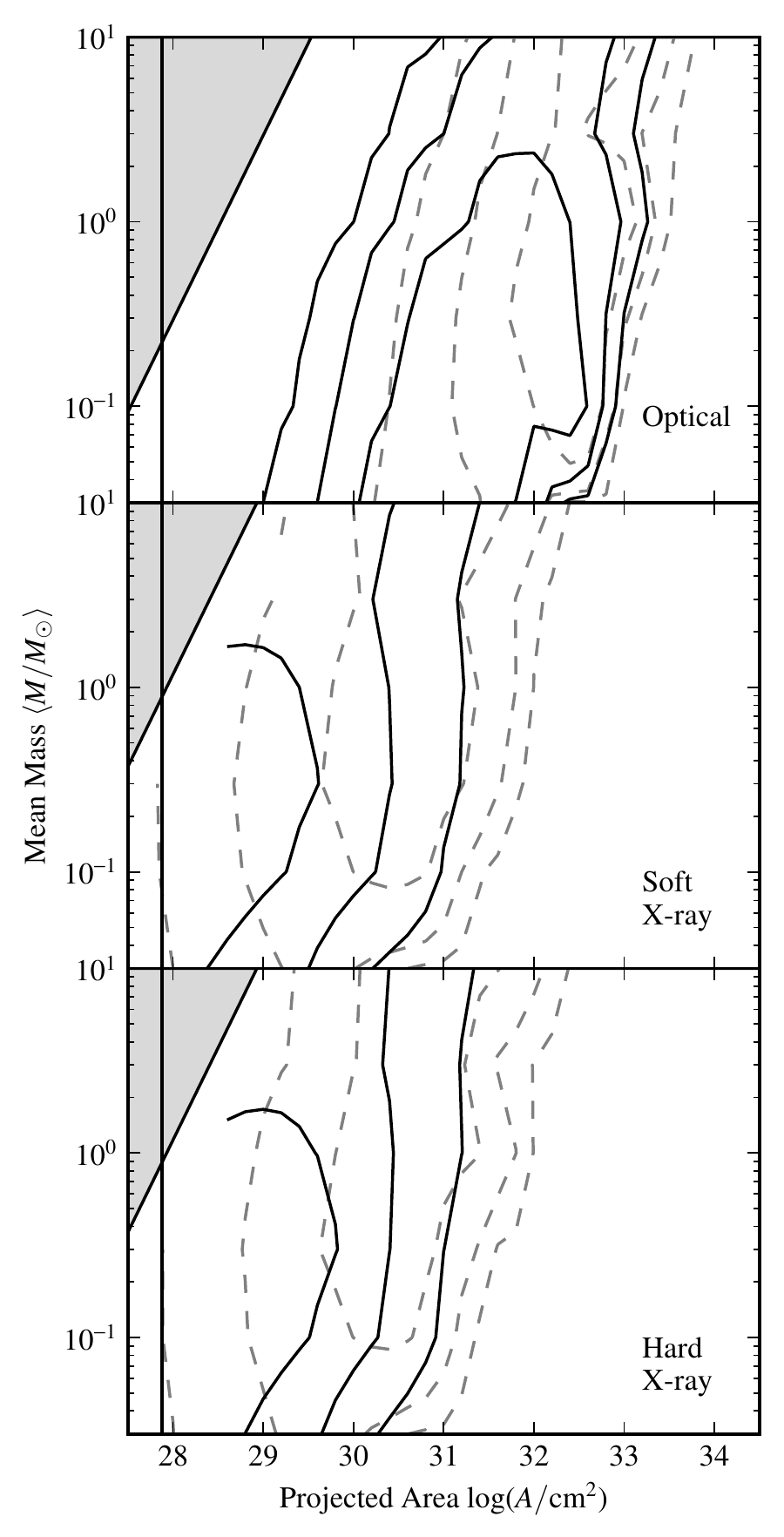}
  \caption{Joint posterior probability distribution for the mean
    microlens mass $\langle M \rangle$ and the projected source area
    $A$ in the $R$ band (top panel), soft X-rays (middle panel), and
    hard X-rays (bottom panel). The solid black (dashed gray) curves
    indicate a logarithmic (linear) prior on the area. The vertical
    line indicates the square of the black hole's gravitational
    radius, as calculated using the black hole mass estimate of
    \protect{\citet{Assef:2011p93}}. The gray region cannot be sampled
    because of the finite resolution of the magnification patterns.}
  \label{fig:massarea}
\end{figure}
\fi

\ifx \emulmacro \undefined
\else
\begin{figure}
  \includegraphics[width=\columnwidth]{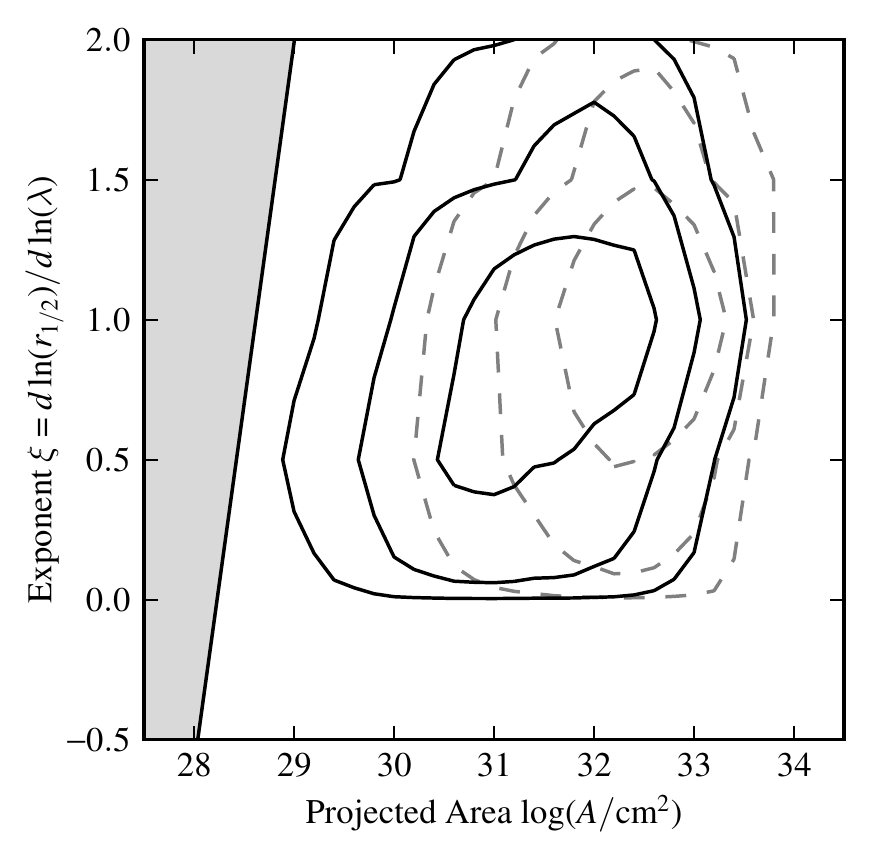}
  \caption{Joint posterior probability distribution for the exponent
  $\xi$ and the projected $R$-band source area $A$. The solid black
  (dashed gray) curves indicate a logarithmic (linear) prior on the
  area. The gray area is excluded because the observed-frame UV disk
  would be smaller than the resolution of the magnification patterns.}
  \label{fig:srcexparea}
\end{figure}
\fi

The soft and hard X-ray half-light radii PDFs are nearly identical,
which is consistent with our expectations given the similarity of
their light curves. To investigate the question of which is larger, we
show in Figure~\ref{fig:sizerat} the posterior PDFs for the logarithm
of the ratios between the sizes in the various bands. The solid curve
indicates the ratio of the hard band size to the soft band size. It
peaks fairly sharply near zero, indicating a ratio of unity. The
symmetric tails on either side mean that we cannot with our current
data distinguish which is the larger, but we can say that their sizes
do not differ too much: at 68\% confidence the logarithm of the ratio
falls between $-0.45$ and $+0.46$. Figure~\ref{fig:sizerat} also
indicates that the X-ray sizes are much smaller than the $R$-band
sizes, with $\Delta \log(r_{1/2}) = -1.40\err{0.44}{0.48}$ for the
soft band and $-1.35\err{0.38}{0.46}$ in the hard band (68\%
confidence). These constraints are tighter than the previous figure
would imply, due to covariances between the optical and X-ray
sizes that are not apparent in Figure~\ref{fig:rhalf}.

Figure~\ref{fig:srcexp} shows the posterior PDF for the power-law
slope of the wavelength dependence of the half-light radius, a
quantity we call $\xi$. For an accretion disk that radiates as a
multi-temperature blackbody, this exponent is the reciprocal of the
power-law slope $\beta$ of the temperature profile $T_\mathrm{eff}(r)
\propto r^{-\beta}$. So for the standard thin disk model, we expect
$\xi = d\log r_{1/2}/d\log \lambda = \beta^{-1} = 4/3$ (neglecting the
effect of the inner disk edge, as is reasonable to do at radii of many
$r_g$). Our probability distribution peaks at a value of 1.0, with a
median value of 0.84, and with 68\% of the probability lying between
values of 0.44 and 1.30. This is consistent with the thin disk value,
though it is interesting that the majority of the probability lies
toward smaller values of $\xi$, or (equivalently) steeper temperature
profiles. Using this posterior distribution, we show in
Figure~\ref{fig:sizerat} the logarithm of the size ratio between the
observed-frame UV and $R$-band disks, and between the $J$-band
and $R$-band disks. These are just scaled versions of the distribution
in Figure~\ref{fig:srcexp}.

Figures~\ref{fig:rhalf} and \ref{fig:sizerat} and the source size
estimates quoted in this section are produced with the mean microlens
mass $\langle M \rangle$ fixed at its most likely value of
$0.3\,M_\odot$. We do this rather than marginalizing over $\langle M
\rangle$ in order to avoid an artificial decrease in the probability
of small source sizes due to the finite resolution of the
magnification patterns. This artificial decrease happens because there
is no contribution to the probability of small sizes from trials with
large $\langle M \rangle$ (and thus with a projected pixel size larger
than the area of the source). To explore the effect of this choice, we
show in Figure~\ref{fig:massarea} the joint posterior PDF for the mean
mass $\langle M \rangle$ and the projected source area $A = \pi
(r_{1/2}/2.44)^2 \cos i$. Some covariance between these parameters can
be seen; this is expected because the area is measured in units of the
square of the microlens Einstien radius. The figure also shows the
resolution limit of the magnification patterns, and it is clear that
the $R$-band source is well-resolved in all cases, but that the X-ray
sources are not much larger than the pixels in the magnification
patterns (despite the higher-resolution patterns we use in these
simulations).

Another concern for the accretion disk simulations is that at large
values of $\xi$ the \textit{HST} UV observations (i.e., the bluest apart
from X-rays) will probe source sizes small enough to be unresolved,
even though the source is resolved in the $R$
band. Figure~\ref{fig:srcexparea} shows the joint posterior PDF of the
exponent $\xi$ and the $R$-band projected source area $A$. We also
show the region where the UV source is unresolved, with $\langle
M \rangle = 0.3 M_\odot$. The probability distribution clearly
converges at small areas, so this concern is unfounded.

\section{Discussion and Conclusions}
\label{sec:discuss}

\heoneone\ is the third lensed quasar to be analyzed using dynamic
microlensing magnification patterns, after
\qtwotwo\ \citep{Poindexter:2010p668, Poindexter:2010p658,
  Mosquera:2013p5009} and
\hefourlong\ \citep{Blackburne:2011p0027}. The inclusion of the random
motions of the stars allows us to constrain the inclination of the
accretion disk, the position angle of its projected major axis, and
the direction of the lens galaxy's motion relative to the quasar. In
this case, the data favor a low disk inclination, with $\cos i = 1.0$
about four times as likely as $\cos i = 0.5$ (see
Figure~\ref{fig:cosi}). This is similar to the result that
\citet{Poindexter:2010p668} find for \qtwotwo, and supports the
``unification'' model for AGNs, which predicts that bright Type 1
quasars such as these have a low inclination. For models with a
nonzero inclination, we find that the major axis of \heoneone's
projected disk is about $150\pm33$ degrees east of north (see
Figure~\ref{fig:diskpa}. We also find that the lens galaxy's motion is
likely toward the northwest or southeast, with a most likely angle of
120 degrees east of north (or 30 degrees west of north), and with 68\%
confidence error bars of 40 degrees. The magnitude of the velocity is
unfortunately not robustly constrained, and depends on our velocity
prior (see Section~\ref{sec:simulations} and Figure~\ref{fig:vel}).

We also produce a posterior PDF for the mean mass of the stars in the
lensing galaxy (see Figure~\ref{fig:meanmass}). The distribution peaks
between $0.1$ and $0.3\,M_\odot$, which is consistent with what is
observed in other lens galaxies \citep{Poindexter:2010p658,
Blackburne:2011p0027} and in the Milky Way \citep{Holtzman:1998p1946,
Zoccali:2000p418}. Our distribution implies a slightly higher mean
mass than that of \citet{Chartas:2009p174}, which is probably due to
differences in the velocity priors. Their analysis uses static
magnification patterns, and (probably more importantly) their peculiar
velocities are smaller than the ones that we have chosen, which would
tend to make their mean mass estimates smaller.

Our results for the half-light radius of the accretion disk in the
observer-frame $R$ band ($0.2\,\mu$m in the rest frame) agree with
those of previous studies of \heoneone\ \citep{Poindexter:2008p34,
  Morgan:2008p80, Chartas:2009p174}. Like these studies, we find that
\heoneone, like several other lensed quasars, has an accretion disk
much larger than would be expected from either the thin disk model or
the quasar flux (see Equations~2 and 3 of
\citet{Poindexter:2008p34}). These estimates, marked ``Flux'' and
``Disk'' respectively, are plotted for the observed-frame $R$ band in
Figure~\ref{fig:rhalf}. \citet{Poindexter:2008p34} suggest a small
value for the accretion disk temperature profile slope $\beta$ as a
possible solution to the discrepancy between the microlensing size and
the flux size, supported by their estimated value of $\beta =
0.61\err{0.21}{0.17}$, smaller than the canonical 0.75. But our result
for the slope of the size-wavelength relation, $\xi = \beta^{-1} =
1.0\err{0.30}{0.46}$, implies a range of 0.77 to 1.85 for
$\beta$. This is roughly consistent with the
\citet{Poindexter:2008p34} result, but some tension remains. It seems
that our new data indicate that the area of the disk changes less with
wavelength than was previously thought. Indeed, our value favors a
temperature profile \emph{steeper} than the standard thin disk
model. This would only serve to exacerbate the flux size discrepancy
\citep[see][]{Morgan:2010p1129}. Multiwavelength observations of more
lensed quasars will shed more light on this ongoing mystery.

Our upper limits on the size of quasar at X-ray wavelengths are quite
strong. It is clear that the majority of the hard and soft X-ray flux
is coming from the innermost $\sim$30\,$r_g$, assuming the
H$\beta$-based black hole mass estimate of
\citet{Assef:2011p93}. This result is comparable to the upper limit
that \citet{Chartas:2009p174} find using a single X-ray data
point. This lack of improvement is partly due to our practice of
dividing the X-ray flux into two bands and of marginalizing over a
variety of source inclinations (both of which broaden the posterior
PDF), and partly due to the fact that we have not yet been able to
sample the X-ray light curve on time scales smaller than the source
crossing time \citep[see][]{Mosquera:2011p96}. For \heoneone\ the soft
and hard X-ray data are very similar, and so we are unable to
determine which band has a larger size, though we do conclude that
their sizes do not differ by more than 0.46\,dex. This result is
similar to those of \citet{Blackburne:2011p0027} for \hefour\ and
\citet{Morgan:2012p52} for \qzeroonelong. The question of the two
X-ray bands' relative sizes is quite interesting, since it can address
the corona/reflection model. If the direct component of the emission
dominates, the hard band, coming from a hotter electron population,
ought to be smaller. But a prominent reflection from the accretion
disk could reverse this result, since the reflected spectrum is harder
than the input spectrum, and may well be more extended. Recent papers
presenting X-ray observations have used simple microlensing models to
address this question \citep{Chartas:2012p137, Chen:2012p24}, but full
microlensing simulations are needed to put rigorous constraints on the
sizes. Some of the X-ray data display fairly strong chromatic
variation, so interesting results should be forthcoming.

\acknowledgements

This research was supported in part by NSF grant AST-1009756. We also
acknowledge the support from the NASA/SAO grants GO0-1112 1A/B/C/D,
GO1- 12139A/B/C, and GO2-13132A/B/C. Support for \textit{HST} programs
\#11732 and \#12324 was provided by NASA through a grant from the
Space Telescope Science Institute, which is operated by the
Association of Universities for Research in Astronomy, Inc., under
NASA contract NAS5-26555. This work was supported in part by an
allocation of computing time from the Ohio Supercomputer Center.

\bibliography{ms}


\ifx \emulmacro \undefined

\newpage

\begin{deluxetable}{cccc}
\tablewidth{0pt}
\tablecaption{SMARTS Light Curves
  \label{tab:smartslc}}
\tablehead{
  \colhead{$\mathrm{HJD}-2450000$} &
  \colhead{A} &
  \colhead{B} &
  \colhead{Filter}
} 
\startdata
$2976.806$ & $+0.11\pm0.01$ & $+1.67\pm0.03$ & $J$  \\
$2976.807$ & $-0.01\pm0.01$ & $+1.38\pm0.02$ & $R$  \\
$2985.815$ & $+0.13\pm0.01$ & $+1.56\pm0.03$ & $J$  \\
$2985.818$ & $-0.02\pm0.01$ & $+1.39\pm0.01$ & $R$  \\
$2993.822$ & $+0.11\pm0.01$ & $+1.53\pm0.02$ & $J$  \\
$2993.826$ & $-0.03\pm0.02$ & $+1.40\pm0.02$ & $R$  \\
$3000.816$ & $+0.10\pm0.01$ & $+1.63\pm0.02$ & $J$  \\
$3000.820$ & $-0.05\pm0.02$ & $+1.40\pm0.02$ & $R$  \\
$3009.812$ & $+0.06\pm0.01$ & $+1.61\pm0.02$ & $J$  \\
$3009.815$ & $-0.09\pm0.01$ & $+1.41\pm0.02$ & $R$
\enddata
\tablecomments{~Light curves are in uncalibrated
magnitudes. This table is published in its entirety online. A portion
is shown here for guidance regarding its form and content.}
\end{deluxetable}

\ifx \emulmacro \undefined
\begin{deluxetable}{cccccc}
\else
\begin{deluxetable*}{cccccc}
\fi
\tablewidth{0pt}
\tablecaption{HST F275W Light Curves
  \label{tab:uvdata}}
\tablehead{
  \colhead{$\mathrm{HJD}-2450000$} &
  \colhead{A} &
  \colhead{B} &
  \colhead{Star a} &
  \colhead{Star b} &
  \colhead{Star c}
} 
\startdata
$5169.817$ & $17.529\pm0.002$ & $18.085\pm0.002$ & $16.094\pm0.001$ & $16.014\pm0.001$ & $18.170\pm0.002$ \\
$5227.008$ & $17.184\pm0.001$ & $18.123\pm0.002$ & $16.074\pm0.001$ & $16.004\pm0.001$ & $18.637\pm0.003$ \\
$5270.614$ & $17.113\pm0.001$ & $18.219\pm0.002$ & \nodata          & \nodata          & $18.831\pm0.003$ \\
$5340.061$ & $17.126\pm0.001$ & $18.589\pm0.003$ & $16.113\pm0.001$ & $16.043\pm0.001$ & $18.344\pm0.002$ \\
$5382.058$ & $16.904\pm0.001$ & $18.327\pm0.002$ & $16.086\pm0.001$ & $16.008\pm0.001$ & $18.223\pm0.002$ \\
$5505.015$ & $17.348\pm0.001$ & $18.328\pm0.002$ & $16.094\pm0.001$ & $16.012\pm0.001$ & $18.345\pm0.002$ \\
$5546.026$ & $17.412\pm0.001$ & $18.298\pm0.002$ & $16.088\pm0.001$ & $16.010\pm0.001$ & $18.267\pm0.002$ \\
$5654.585$ & $17.242\pm0.001$ & $18.023\pm0.002$ & $16.107\pm0.001$ & $16.026\pm0.001$ & $18.397\pm0.002$ \\
$5729.072$ & $17.207\pm0.001$ & $18.258\pm0.002$ & $16.113\pm0.001$ & $16.048\pm0.001$ & $18.645\pm0.003$ \\
$5877.216$ & $17.147\pm0.001$ & $18.047\pm0.002$ & $16.090\pm0.001$ & $16.013\pm0.001$ & $18.219\pm0.002$
\enddata
\tablecomments{~Light curves are in ST magnitudes. In this filter, the
offset from ST to AB magnitudes is $m_\mathrm{AB}-m_\mathrm{ST} =
1.532$.}
\ifx \emulmacro \undefined
\end{deluxetable}
\else
\end{deluxetable*}
\fi

\newpage

\begin{figure}
  \centering
  \includegraphics[width=\textwidth]{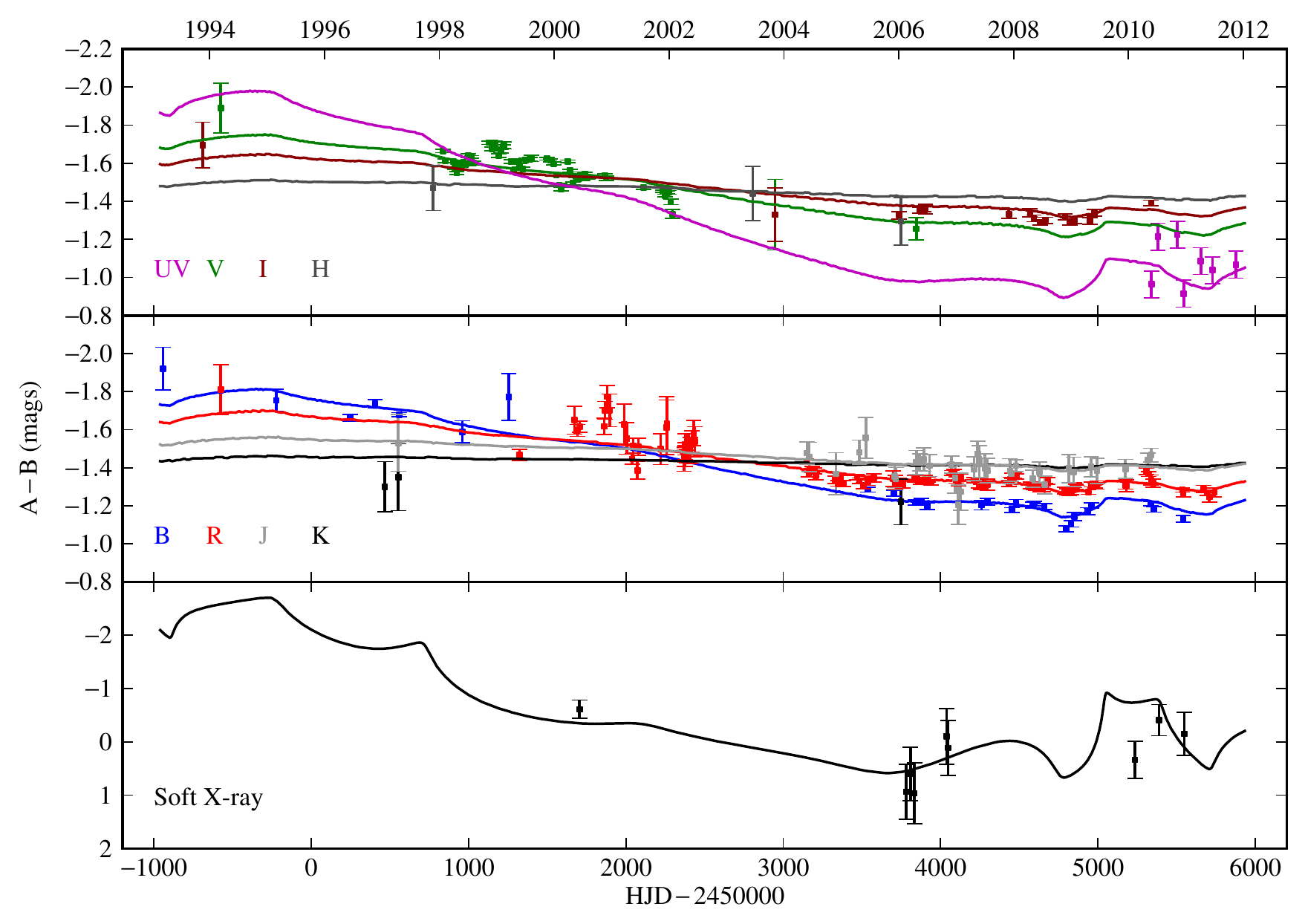}
  \caption{Time delay-corrected difference in magnitudes between the A
    and B images of \heoneone\ in each filter as a function of time
    (points with error bars). Some systematic errors have been added
    as described in Section~\protect{\ref{sec:data}}. Curves show one
    of the best-fitting out of millions of microlensing light curves
    found by our simulation. The soft X-ray light curve is in its own
    panel because of its much larger range. The hard X-ray data are
    nearly identical, and we do not show them.}
  \label{fig:lc}
\end{figure}

\begin{figure}
  \centering
  \includegraphics[width=\columnwidth]{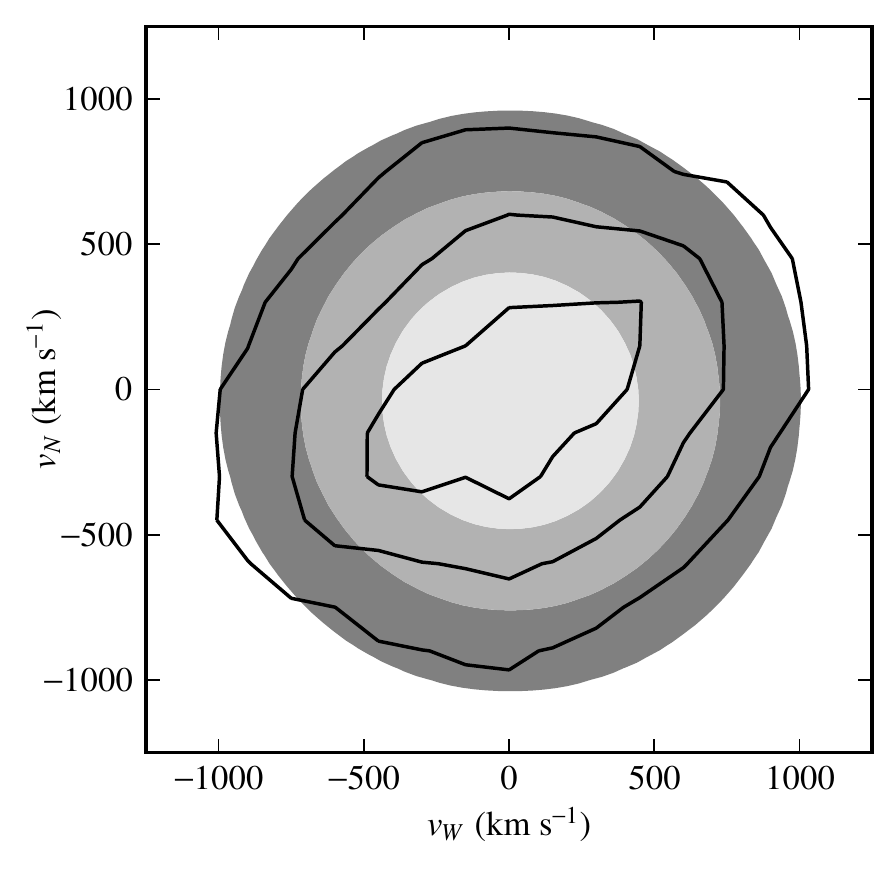}
  \caption{Filled contours indicate the 68\%, 95\%, and 99\% confidence
    levels of the velocity prior described in
    Section~\ref{sec:simulations}. Open contours show the posterior
    probability distribution of the transverse lens velocity (relative
    to that of the source and the observer). The direction of motion
    of the lens galaxy is $120\pm40$ degrees, with a 180-degree
    degeneracy due to the symmetry of the elliptical disk model.}
  \label{fig:vel}
\end{figure}

\begin{figure}
  \centering
  \includegraphics[width=\columnwidth]{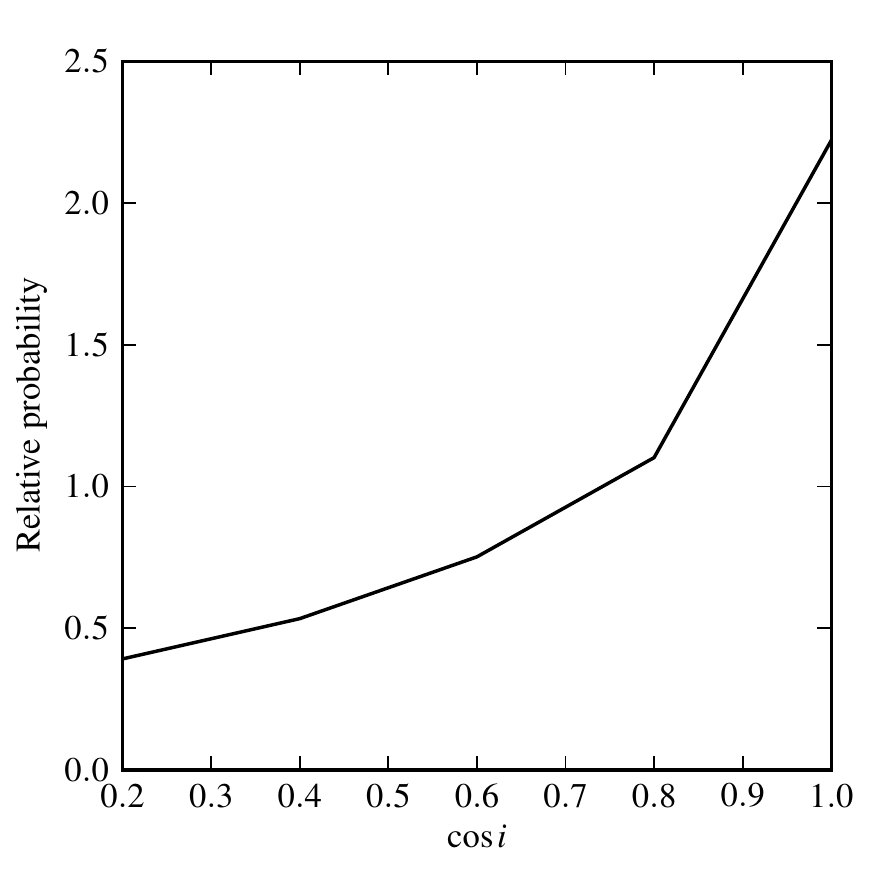}
  \caption{Posterior distribution for the accretion disk inclination
    $\cos i$. A face-on disk has $\cos i = 1$.}
  \label{fig:cosi}
\end{figure}

\begin{figure}
  \centering
  \includegraphics[width=\columnwidth]{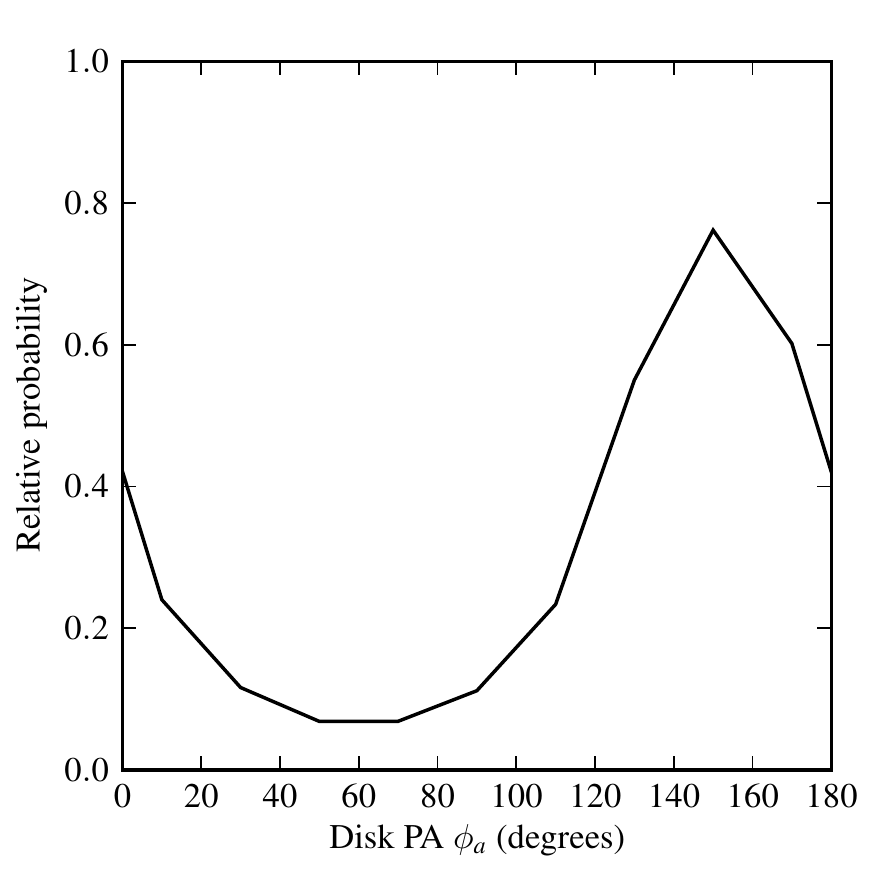}
  \caption{Posterior distribution for the accretion disk major axis
    position angle $\phi_a$, measured in degrees East of
    North. Face-on ($\cos i = 1$) solutions are excluded in
    calculating this result.}
  \label{fig:diskpa}
\end{figure}

\begin{figure}
  \centering
  \includegraphics[width=\columnwidth]{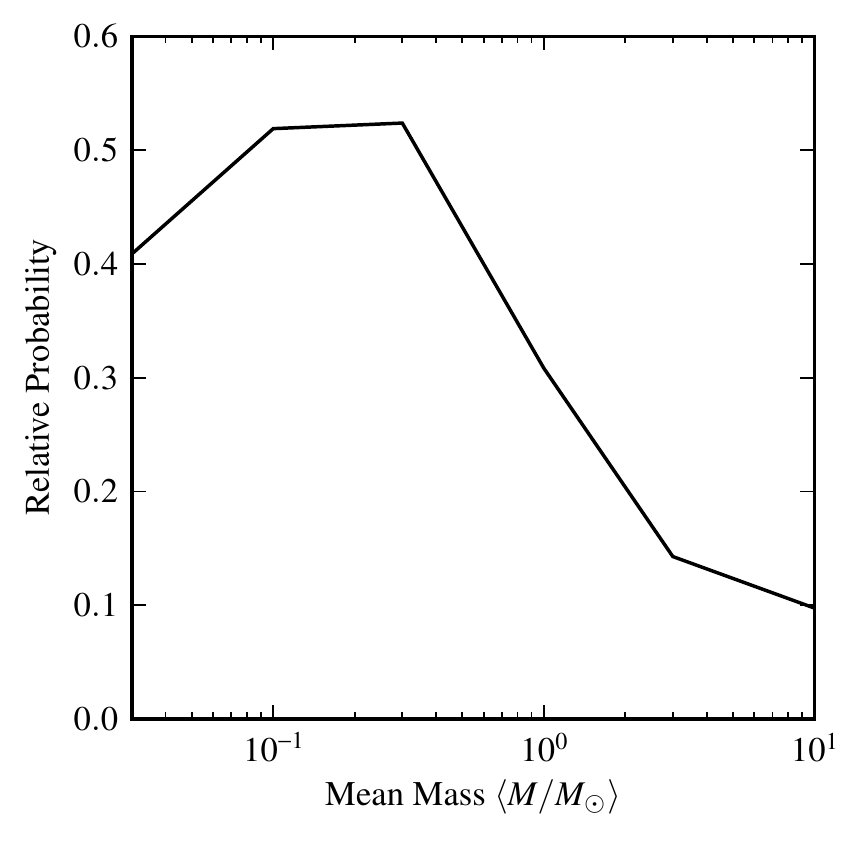}
  \caption{Posterior distribution for the mean mass of the stars in
    the lensing galaxy.}
  \label{fig:meanmass}
\end{figure}

\begin{figure}
  \centering
  \includegraphics[width=0.7\columnwidth]{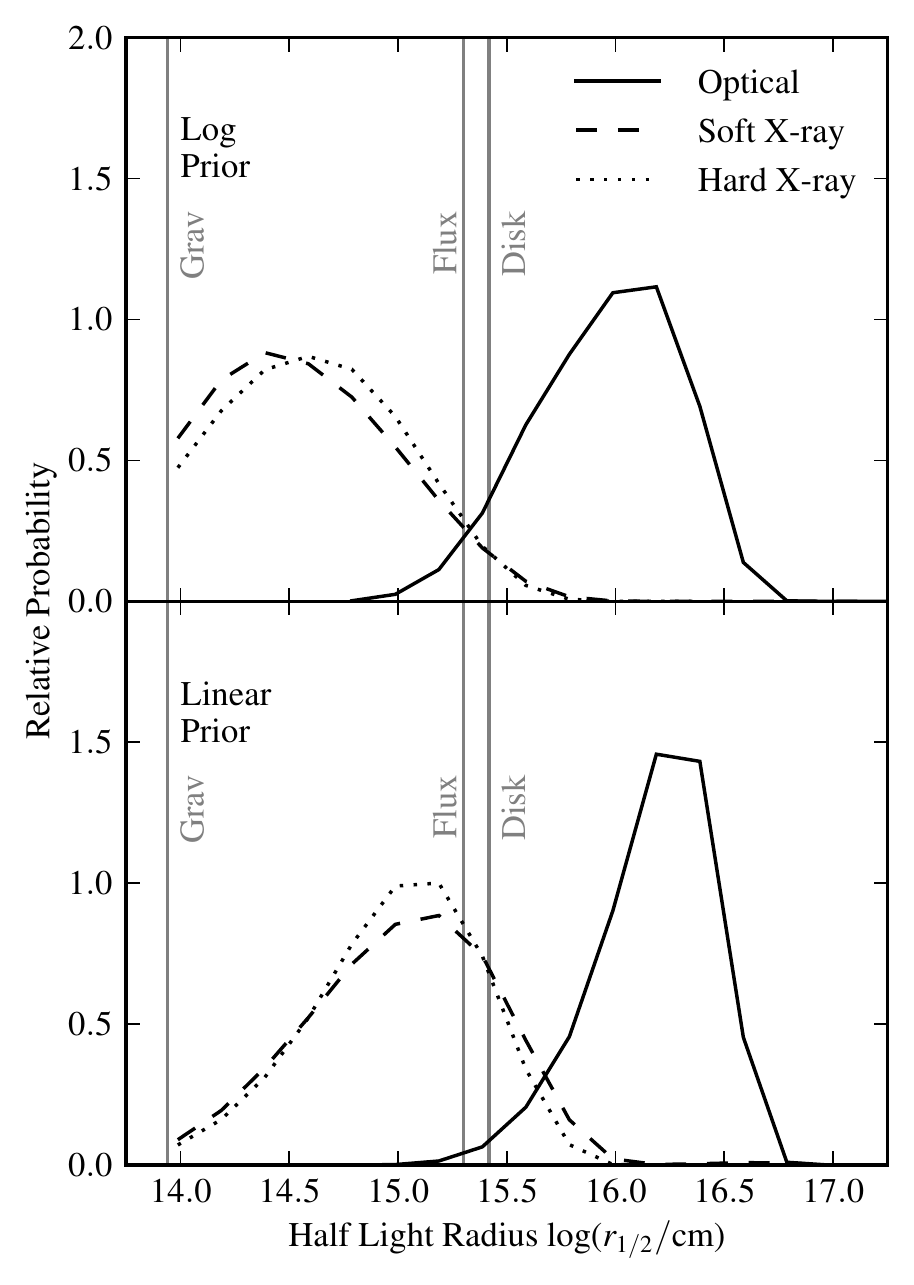}
  \caption{Posterior distribution for the half-light radius of the
    quasar accretion disk in the $R$ band and the two X-ray bands. In
    the upper panel we use a logarithmic prior, and in the lower panel
    we use a linear prior. The stellar mean mass $\langle M/M_\odot
    \rangle$ is fixed at 0.3. The gravitational radius of the black
    hole and the $R$-band size estimates based on the quasar flux and
    the thin disk model are shown as a vertical lines, marked
    ``Grav,'' ``Flux,'' and ``Disk'' respectively. The ``Grav'' and
    ``Disk'' values assume the H$\beta$ black hole mass estimate of
    \protect{\citet{Assef:2011p93}}, and can shift up to
    $\sim$0.5\,dex based on the uncertainties or by using the
    H$\alpha$ or \textsc{Civ} estimates.}
  \label{fig:rhalf}
\end{figure}

\begin{figure}
  \centering
  \includegraphics[width=\columnwidth]{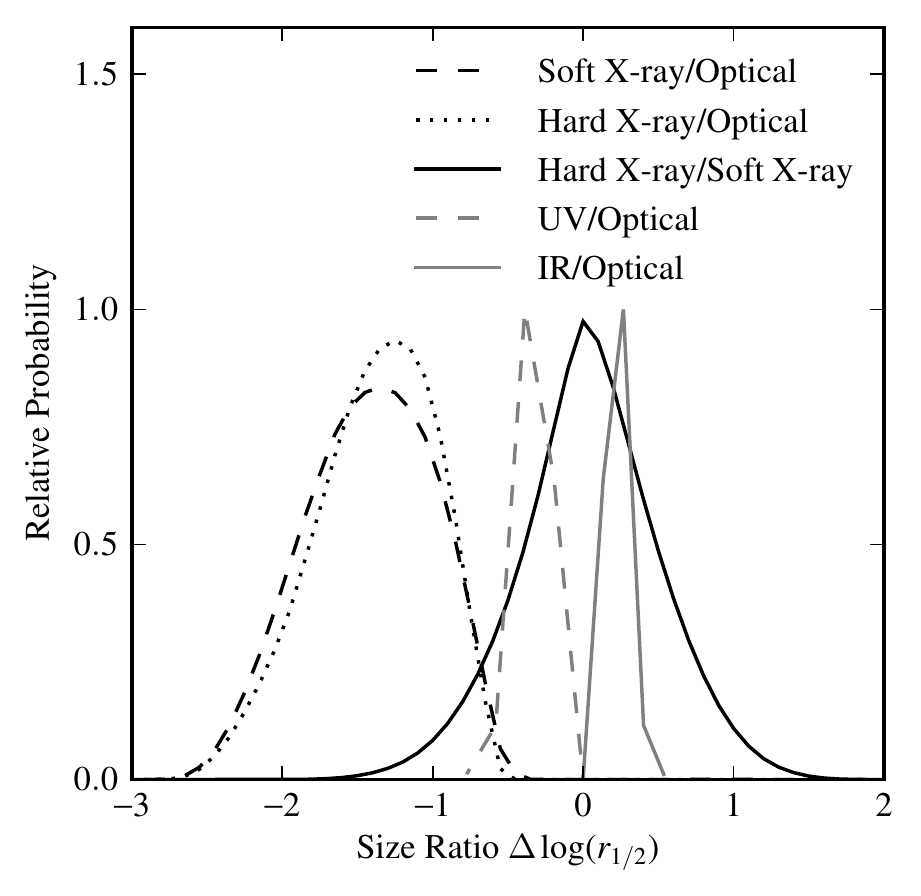}
  \caption{Posterior distributions for the ratios of the quasar's
    half-light radius. The black curves indicate ratios of the X-ray
    sizes; for these curves the mean mass $\langle M \rangle$ is set
    to $0.3 M_\odot$. The X-ray emission is much more compact than the
    observed-frame $R$-band emission (labeled ``Optical''). Also, our
    X-ray data rule out a very large difference in the sizes of the
    hard and soft X-ray sources. The gray curves indicate the size of
    the accretion disk at the (observed-frame) UV and $J$-band
    wavelengths, relative to the $R$-band size. Since we parameterize
    the disk using the $R$-band size and the wavelength slope $\xi$,
    the gray curves are simply scaled versions of the posterior
    distribution for $\xi$ (see \protect{Figure~\ref{fig:srcexp}}).}
  \label{fig:sizerat}
\end{figure}

\begin{figure}
  \centering
  \includegraphics[width=\columnwidth]{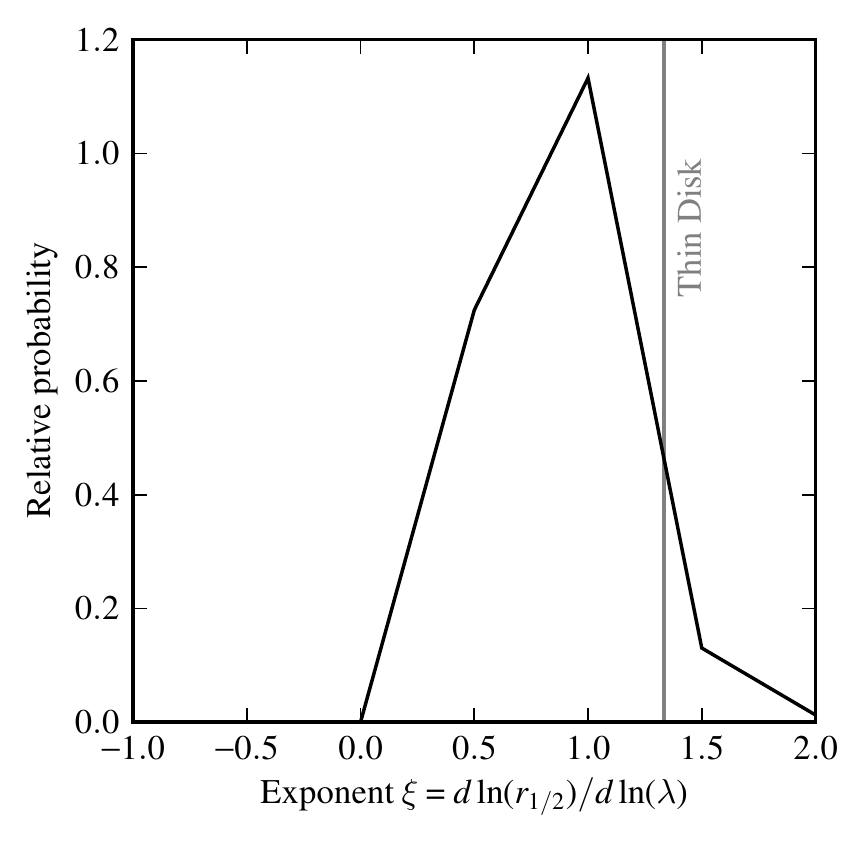}
  \caption{Posterior probability distribution for the power-law slope
    of the wavelength dependence of the half-light radius. The
    standard thin disk model predicts a value of 4/3 since
    $T_\mathrm{eff} \propto r^{-3/4}$.}
  \label{fig:srcexp}
\end{figure}

\begin{figure}
  \centering
  \includegraphics[width=0.5\columnwidth]{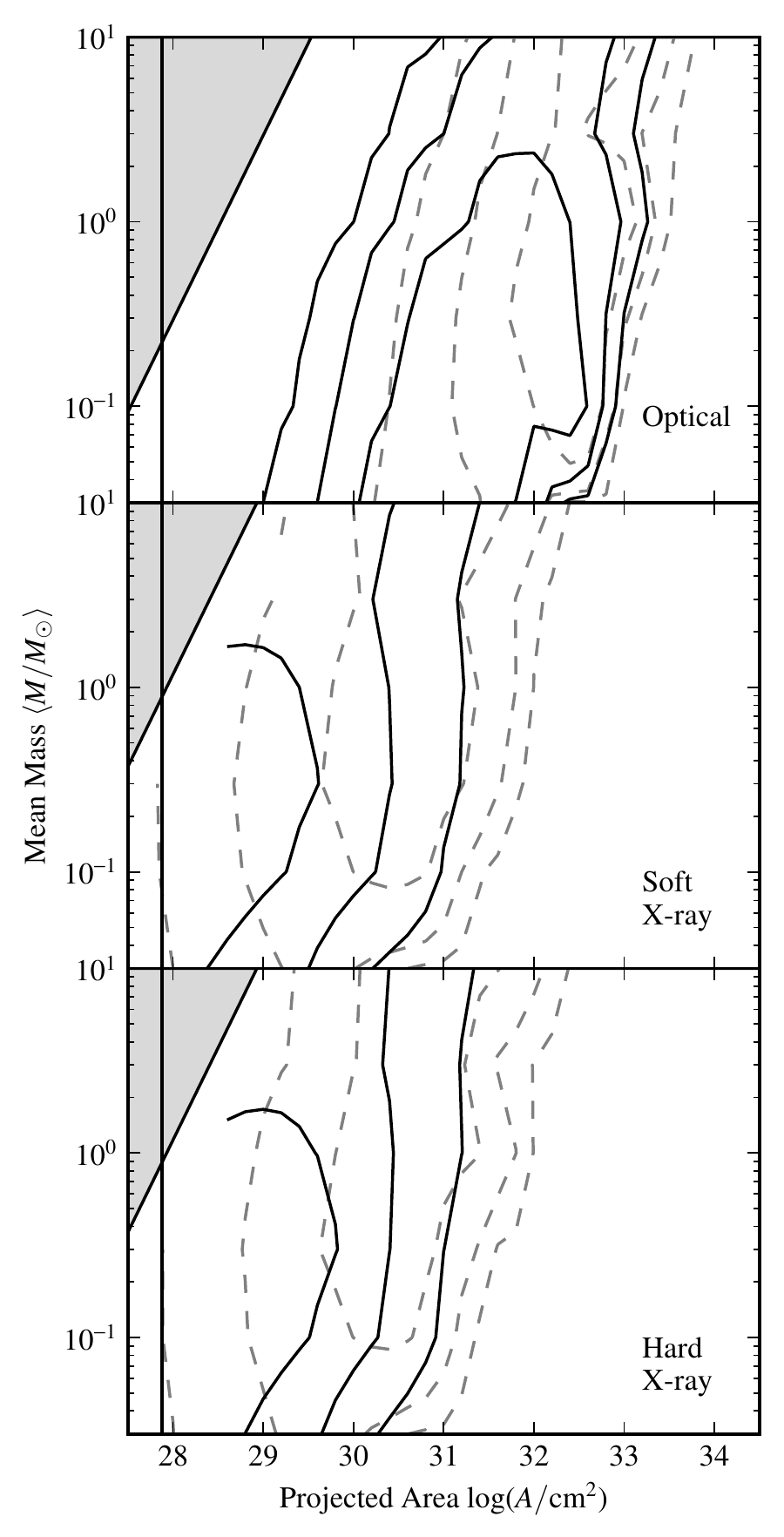}
  \caption{Joint posterior probability distribution for the mean
    microlens mass $\langle M \rangle$ and the projected source area
    $A$ in the $R$ band (top panel), soft X-rays (middle panel), and
    hard X-rays (bottom panel). The solid black (dashed gray) curves
    indicate a logarithmic (linear) prior on the area. The vertical
    line indicates the square of the black hole's gravitational
    radius, as calculated using the black hole mass estimate of
    \protect{\citet{Assef:2011p93}}. The gray region cannot be sampled
    because of the finite resolution of the magnification patterns.}
  \label{fig:massarea}
\end{figure}

\begin{figure}
  \centering
  \includegraphics[width=\columnwidth]{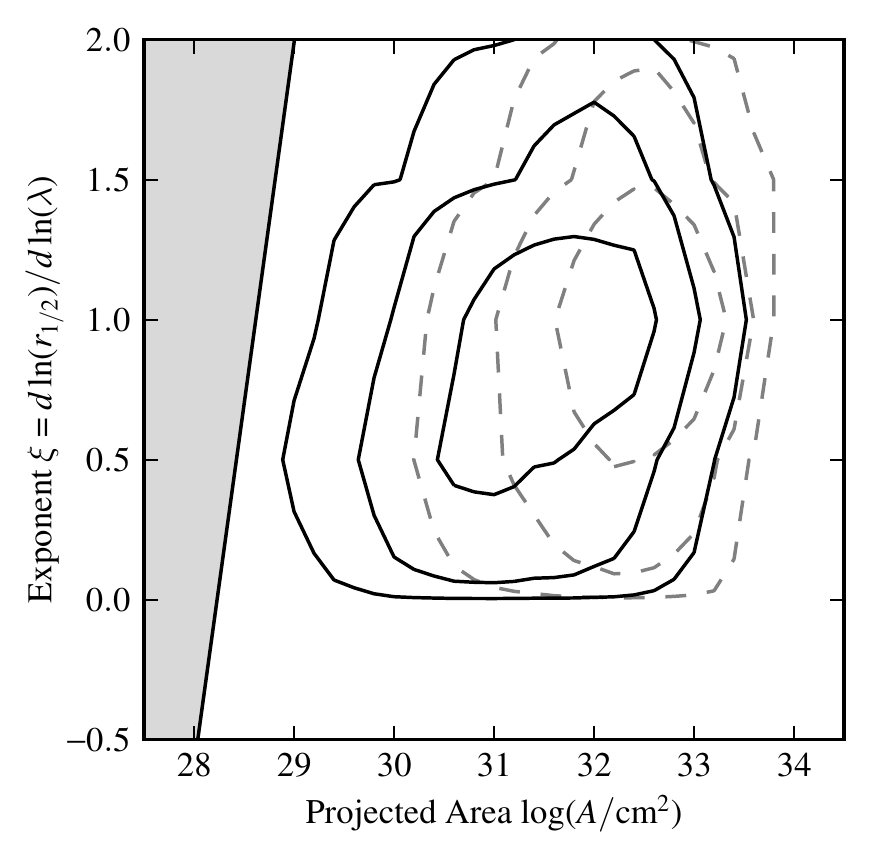}
  \caption{Joint posterior probability distribution for the exponent
  $\xi$ and the projected $R$-band source area $A$. The solid black
  (dashed gray) curves indicate a logarithmic (linear) prior on the
  area. The gray area is excluded because the observed-frame UV disk
  would be smaller than the resolution of the magnification patterns.
}
  \label{fig:srcexparea}
\end{figure}

\else
\fi

\end{document}